\documentclass[12pt,preprint]{emulateapj} % single-spaced, one column

\usepackage{lscape}
\usepackage{graphicx}
\usepackage{natbib}
\usepackage{endnotes}
\usepackage{subfigure}
\nocite{*}

\newcommand{\etal}{et~al.\/}

\newcommand{\Hline}[1]{\mbox{H{\footnotesize {#1}}}}
\newcommand{\Halpha}{\Hline{\mbox{$\alpha$}}}

\newcommand{\HI}{{\sc Hi}}

\newcommand{\HIPASS}{{\sc HiPASS}}

\newcommand{\ha}{H${\alpha}$}
\shortauthors{Werk \etal}
\shorttitle{Outlying HII Regions}

\begin{document} 
\slugcomment{Accepted by AJ: November 2009}

\title{Outlying HII Regions in HI-Selected Galaxies}
                 
\author{J. K.\ Werk\altaffilmark{1,2},
M. E.\ Putman\altaffilmark{2},
G. R.\ Meurer\altaffilmark{3},
E. V.\ Ryan-Weber\altaffilmark{4}, 
C. Kehrig\altaffilmark{1}, D.A.\
  Thilker\altaffilmark{3}, J.\ Bland-Hawthorn\altaffilmark{5}, M.J.\
  Drinkwater\altaffilmark{6},R.C.\ Kennicutt, Jr.\altaffilmark{7}, O.I.\
  Wong\altaffilmark{8}, K.C.\ Freeman\altaffilmark{9}, M.S.\ Oey\altaffilmark{1}, M.A.\ Dopita\altaffilmark{9}, 
  M.T.\ Doyle\altaffilmark{10}, H.C.\ Ferguson\altaffilmark{11}, D.J.\ Hanish\altaffilmark{1},  T.M.\ Heckman\altaffilmark{3},
  V.A.\ Kilborn\altaffilmark{4}, J.H.\ Kim\altaffilmark{12}, P.M.\ Knezek\altaffilmark{13}, B.\ Koribalski\altaffilmark{14},
   M.\ Meyer\altaffilmark{15}, R.C.\ Smith\altaffilmark{16}, M.A.\ Zwaan\altaffilmark{17}
}

\altaffiltext{1}{Department of Astronomy, University of Michigan, 500 Church St., 
		Ann Arbor, MI 48109, $jwerk@umich.edu$}
		\altaffiltext{2}{Department of Astronomy, Columbia University, 550 West
                  120th Street, New York, NY 10027}

\altaffiltext{3}{Department of Physics and Astronomy, The Johns Hopkins University, 
		Baltimore, MD 21218-2686}
\altaffiltext{4}{Centre for Astrophysics and Supercomputing, Swinburne University of Technology, Mail H39, PO Box 218, Hawthorn, VIC 3122, Australia}
\altaffiltext{5}{Institute of Astronomy, School of Physics, University
                 of Sydney, Australia}

                   \altaffiltext{6}{Department of Physics, University of Queensland, 
                  Brisbane, QLD 4072, Australia}
                  \altaffiltext{7}{Institute of Astronomy, University of Cambridge, Madingley Road, Cambridge, 
		CB3 0HA, UK}
                 \altaffiltext{8}{Department of Astronomy, Yale University, New Haven, 
                 CT 06520}
\altaffiltext{9}{Research School of Astronomy and Astrophysics, 
                  Australian National University, Cotter Road, Weston 
                  Creek, ACT 2611, Australia}
\altaffiltext{10}{Department of Physics, University of Queensland, 
                  Brisbane, QLD 4072, Australia}
\altaffiltext{11}{Space Telescope Science Institute, 3700 San Martin
                 Drive, Baltimore, MD 21218}
                 \altaffiltext{12} {Center for the Exploration of the Origin of the Universe, Department of Physics and Astronomy, Seoul National University, Seoul, Korea}
                  \altaffiltext{13}{WIYN Consortium, Inc., 950 North Cherry Avenue, Tucson,
                 AZ 85726}

                 \altaffiltext{14}{Australia Telescope National Facility, CSIRO, P.O. Box
                 76, Epping, NSW 1710, Australia}
                                 	\altaffiltext{15}{School of Physics, The University of Western Australia, Crawley 6009}
\altaffiltext{16}{Cerro Tololo Inter-American Observatory (CTIO), Casilla
                 603, La Serena, Chile} 
                 \altaffiltext{17}{European Southern Observatory,
                 Karl-Schwarzschild-Str.\ 2, 85748 Garching b.\ 
                 M\"{u}nchen, Germany}

\begin{abstract}

We present results from the first systematic search for outlying HII regions, as part of a sample of 96  emission-line point sources (referred to as ELdots - emission-line dots) derived from the NOAO Survey for Ionization in Neutral Gas Galaxies (SINGG). Our automated ELdot-finder searches SINGG narrow-band and continuum images for high equivalent width point sources outside the optical radius of the target galaxy ($>$ 2 $\times$ r$_{25}$ in the R-band). Follow-up longslit spectroscopy and deep  GALEX images (exposure time $>$ 1000 s) distinguish outlying HII regions from background galaxies whose strong emission lines ([OIII], H$\beta$ or [OII]) have been redshifted into the SINGG bandpass.  We find that these deep GALEX images can serve as a substitute for spectroscopic follow-up because outlying HII regions separate cleanly from background galaxies in color-color space. We identify seven SINGG systems with outlying massive star formation that span a large range in H$\alpha$ luminosities corresponding to a few O stars in the most nearby cases, and unresolved dwarf satellite companion galaxies in the most distant cases. Six of these seven systems feature galaxies with nearby companions or interacting galaxies. Furthermore, our results indicate that some outlying HII regions are linked to the extended-UV disks discovered by GALEX, representing emission from the most massive O stars among a more abundant population of lower mass (or older) star clusters.  The overall frequency of  outlying HII regions in this sample of gas-rich galaxies is 8 - 11\% when we correct for background emission-line galaxy contamination ($\sim75$\% of ELdots).

\end{abstract}


\section{Introduction}
\label{intro}
In the last 10 years, deep H$\alpha$ imaging has confirmed the occurrence of recent star formation far beyond the main optical bodies of  galaxies, in both discrete knots with no underlying continuum emission and faint outer spiral arms (eg. \nocite{ferguson98, emma,bfq97} Bland-Hawthorn, Freeman, \& Quinn 1997; Ferguson et al. 1998; Ryan-Weber et al. 2004).  More recently, the {\it{Galaxy Evolution Explorer}} (GALEX) satellite has broadened our picture of outer-galaxy star formation by revealing that over $\sim30$\% of spiral galaxies possess UV-bright extensions of their optical disks, implying that low-intensity star formation in the outskirts of galaxies is not particularly rare \citep{Thilker07}.  This mode of star formation in the outer disks of galaxies may be a natural extension of inside-out disk formation, in which galactic stellar disks grow gradually with time following the condensation and assembly of their gaseous disks (\nocite{whiteandfrenk91, bouwens97} e.g. White \& Frenk 1991; Bouwens, Cayon, \& Silk 1997).  Moreover, extended-UV (XUV) emission and outlying HII regions sometimes are associated with previous or ongoing galaxy interactions \citep{Thilker07, werk08}.  Probing the nature and frequency of outer-galaxy star formation can facilitate our understanding of the disk building process, and the ways in which new stellar populations emerge via galaxy interactions.  Star formation in galactic outskirts also has relevance to discussions involving star formation gas density thresholds, as it often occurs in a low-density environment (e.g. \nocite{gerhard,mendes04, emma}  Gerhard et al. 2002; Mendes de Oliveira et al. 2004; Ryan-Weber et al. 2004). One way to increase our understanding of star formation ``thresholds" is to assemble a sample of massive stars forming in low column-density gas outside the influence of the existing stellar population in a galactic disk.

The \ha~ and UV radiation in the extended disks trace different ranges of stellar mass: UV radiation follows primarily the O and B stars, while  \ha~ emission traces just the O stars. Studies of spiral galaxies with XUV disks \citep{thilker05, gildepaz05} have indicated that the spatial extent of star formation in outer disks is underestimated by looking for HII regions alone (via \ha~ imaging), as \ha~emission tends to be even less widespread than predicted by typical stellar population synthesis models that incorporate a power-law IMF with a slope of $\alpha\sim2.35$. The reasons for this relative lack of \ha~emission at large galactocentric radii could be many, including but not limited to stochastic fluctuations in the IMF at low stellar luminosities \citep{boissier06}, a top-light IMF in the remotest reaches of galaxies \citep{meurer09}, or a large fraction of escaping ionizing photons from the less dense outer-galactic regions.  The extent of this underestimation, specifically at large projected galactocentric distances, remains unquantified, as no comprehensive systematic study of \ha~emission in the outer reaches of galaxies has yet emerged.

Here, we return to  \ha~imaging to find star formation in the outskirts of galaxies. As a star formation tracer that is sensitive to  only the highest mass stars, \ha~provides an important complement to the recent GALEX results.  We perform an automated search to find outlying compact sources of net line emission using the imaging data from the Survey for Ionization in Neutral Gas Galaxies (SINGG; Meurer et al 2006; hereafter M06\nocite{SINGG}). The large angular area outside the optical radius of galaxies in each 14.7\'~SINGG field presents the opportunity to search for outlying HII regions at large radii and perform a blind search for background emission-line sources. We initially refer to both types of  sources as ``ELdots" for their appearance as emission line dots in the images.  In all cases, ELdots exhibit strong emission lines in the SINGG narrowband filter, and are point sources well outside the broadband optical emission of nearby galaxies (r$>$2 $\times$ r$_{25}$). ELdots can be outlying \ha-emitting HII regions at a similar velocity to the HIPASS source galaxy or background  galaxies emitting a different line ([OIII], [OII], or H$\beta$) that is redshifted into the narrow filter passband used for the observations. To distinguish between these options, we present follow-up spectroscopy and deep archival GALEX images for a subsample of ELdots.

 In this study, we are primarily interested in the \ha-emitting ELdots, although we tabulate the properties of the background galaxies as well. While a number of previous works have discovered ``intergalactic" HII regions, this is the first systematic search of an unbiased sample of gas rich galaxies for such emission. Furthermore, although many studies have chosen ``intergalactic" as a modifier (see \nocite{boquien09} Boquien et al. 2009 for a detailed description of the terminology), we opt instead for ``outlying." These HII regions are distinct from the central galactic star formation, and more sparse, yet they often lie in extended neutral gas and/or an extended UV component associated with the galaxy. Outlying HII regions (abbreviated outer-HIIs) provide a unique laboratory to understand star formation under relatively extreme conditions (e.g. low neutral gas column density, weak galaxy potential), and may shed light on the full extent of stellar disks \citep{blandhawthorn05, irwin05}.

The paper proceeds as follows: in Section 2, we describe the SINGG sample; in Section 3, we describe our ELdot sample selection; in Section 4, we present spectroscopic observations of a subsample of ELdots; in Section 5 we examine the ELdots in deep GALEX images, where available from the archives; in Section 6, we discuss the properties of our ELdot sample;  in Section 7, we discuss extended star formation in SINGG; and in Section 8 we summarize and present the key findings of this study.

\section{SINGG data}
We searched for ELdots in the SINGG Release 1 (SR1) imaging data
presented by M06.  A full description of the sample, observations, and
data reduction techniques used by SINGG is presented by M06.  Here we
summarize the most relevant facts and refer the interested reader to M06
for details.  The SINGG sample was selected from the HI Parkes  All-Sky Survey (HIPASS; \nocite{barnes01} Barnes et al. 2001) purely on the
basis of \HI\ mass and recessional velocity.  SR1 consists of
observations of 93 \HIPASS\ targets, which contain a total of 111
emission line galaxies.  We refer to the  \HIPASS\ targets/SINGG fields as JXXXX-YY, where XXXX is RA in hours and minutes and YY is declination in degrees. In the text, we refer to the galaxy(ies) contained in the field with their common optical names (e.g. NGC 1512).  For reference, table 6 of M06 contains the common optical name for each HIPASS target. 

The SINGG images obtained with the CTIO
1.5m telescope have a field of view of $14.7'$ and typically consisted
of three dithered images with a combined exposure time of 1800s in  a narrow band filter (band pass widths $\sim 30
- 75$\AA) to isolate the redshifted \Halpha\ emission, and 360s exposures
in the $R$ band for the continuum.  Following standard CCD image
processing, a net \Halpha\ image was created from differencing the
combined images in each filter after a suitable convolution to match
point spread functions.  Observations of spectro-photometric standards
were used to flux calibrate the data.  The median seeing was 1.6$''$ while the median
5$\sigma$ limiting \Halpha\ flux for a point source is $2.6\times
10^{-16}\, {\rm erg\, cm^{-2}\, s^{-1}}$.  The median Hubble flow distance of the entire SINGG sample is 18.5 Mpc, corresponding to a median FOV of 79 kpc. Foreground Galactic dust
extinction was removed using the reddening maps of \citet{schlegel98}.  All magnitudes are in the AB magnitude system.

The discovery of outer-HIIs in SINGG was initially presented by \nocite{emma} Ryan-Weber et al. (2004; hereafter RW04), who performed a by-eye search of the SR1 data for ELdots. This work represents a continuation of RW04, with the addition of an automated search algorithm developed specifically to pick out ELdots, follow-up spectroscopy, and comparison with existing deep UV GALEX data. The initial sample presented in RW04 is augmented extensively by our more systematic search and fluxes are updated to include the more recent calibration presented in M06. 

\section{Sample of ELdots}
\label{ELdotfinder}
\subsection{Selection Criteria}

We approached our sample selection with the goal of finding HII regions in the outskirts of galaxies or the IGM. We have automated the search for ELdots  in an IDL program written specifically for the purpose of finding compact, high equivalent width sources at distances greater than 2 $\times$ r$_{25}$ from the nearest SINGG galaxy. The program incorporates SExtractor \citep{bertin} for object detection and photometry, and includes the following steps: 
\begin{figure}[h!]
\centering
\mbox{\subfigure[]{\includegraphics[width=2.5in]{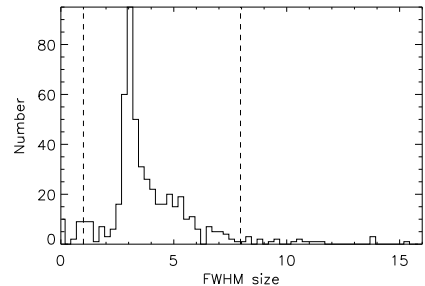}}
}
\mbox{\subfigure[] {\includegraphics[width=2.5in]{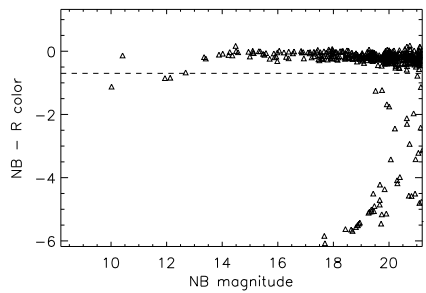}} 
}
\mbox{\subfigure[] {\includegraphics[width=2.5in]{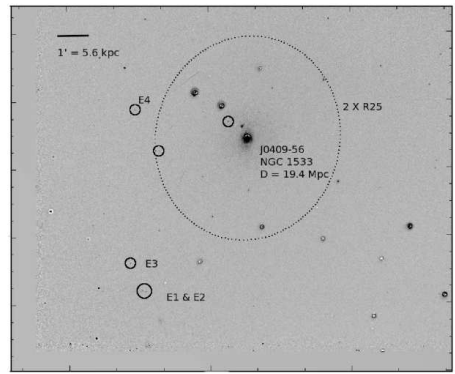}} 
}

\caption[]{{{(a): The distribution of objects' FWHM sizes (in pixels) in the same field, J0409-56, as determined by SExtractor. Dashed vertical lines show the minimum and maximum size thresholds for an object to be classified as an ELdot.  (b): The distribution of objects' raw SExtractor NB$-$R colors in the SINGG field J0409-56. The dashed line represents the color limit of -0.7. ELdots are selected to have colors less than this value. The diagonal line of objects below the color limit represents those objects which have no R-band emission, and are all given the same default upper-limit R-band magnitude. (c): Continuum-subtracted \ha~ image of J0409-56, with the ELdots circled and labeled. The dotted line represents the 2$\times$r$_{25}$ elliptical isophote determined from the SINGG R-band image. The two circled sources within 2$\times$r$_{25}$ are ultimately rejected by the finder, despite meeting size and color criteria. We give the angular and equivalent physical scale in the top left corner. The projected galactocentric distances of the ELdots in this field range from 19 kpc (E4) to 31 kpc (E1 and E2).}}}
\label{diag}
\end{figure}

\begin{itemize} 

\item The object detection threshold used by the SExtractor requires that at least three adjacent pixels have counts $3.0\sigma$ above the image background level. We use 32 deblending sub-thresholds, with a minimum contrast parameter of 0.005, assuring the detection of small, unresolved objects. We determine magnitude zero-points from the SINGG image photometry described in M06. Since SExtractor is run on sky-subtracted narrow-band and continuum images, we perform no additional background subtraction. Minor sky gradients in a few of the SINGG images amount to variations at the $0.1\sigma$ level in the worst cases, and employing a background subtraction in these cases makes virtually no difference in the final sextractor object catalogs.  

\item To limit the number of spurious detections in the automatically-produced candidate list, we include only those objects which lie in areas of the SINGG images composed of at least two exposures. Usually this amounts to rejecting approximately 100 pixel-wide strips on each edge of the image. 

\item To keep cosmic rays, hot pixels, and larger angular sized galaxies out of our final sample of ELdots, we implement two cuts in FWHM, calculated individually for each frame.  The FWHM size of the object must be greater than 1 pixel and less than the mean image FWHM $+$ 3$\sigma$. We show an example of this criterion, and the distribution of object FWHM sizes in the SINGG field J0409-56, in Figure1a. The mean FWHM of objects in the SINGG images is typically near 4 pixels ($\sim1.6$\arcsec~ at the pixel scale of SINGG) while the standard deviation ($\sigma$) of this size distribution ranges between 1.5 and 2 pixels. Thus, objects with angular sizes greater than $\sim4$\arcsec~ (8$-$10 pixels) are eliminated from the sample of ELdots. This maximum angular size represents a broad range of physical sizes when we consider the range of distance spanned by the SINGG sample. The minimum and maximum distances of SINGG targets, roughly 4 Mpc and 70 Mpc,  yields maximum HII region sizes of 80 pc to 1400 pc, respectively. We discuss this size selection effect, along with other caveats related to the SINGG ELdots in Section \ref{selectioneffects}.  

\item To ensure we include only those objects with real emission lines and little or no continuum emission, we use a NB$-$R (NB$=$ narrow band; R$=$ R band) color cut of $-0.7$. This cut in color roughly corresponds to a cut in equivalent width of 20 \AA. Depending on the filter transmission curve, it can be as low as 5 \AA~ and as high as 35 \AA. Figure 1b plots the raw NB$-$R color versus NB magnitude for all of the objects meeting the above criteria in the field J0409-56, and shows this color limit. 

\item The program uses the surface-brightness profiles of each SINGG galaxy in an image, and rejects any object that lies within two times the $\mu$$_{R} = 25$ mag arcsec$^{-2}$ elliptical isophote of any of the potential host galaxies in the image.  r$_{25}$  is a standard galactocentric distance that generally denotes the full extent of a galaxy's optical disk. Traditionally, it is defined in the B-band. Here, we use 2 $\times$ r$_{25}$ in the R-band as a threshold to distinguish between HII regions in the disk and those outside the optical disk.  Using values tabulated by the NASA Extragalactic Database (NED), we find the B-band r$_{25}$ values for the SINGG galaxies (where available) range between 1.5 and 2.2 $\times$ r$_{25}$ in the R-band, with an average at 1.9 $\times$ r$_{25}$ in the R-band.  We show an example of our 2 $\times$ r$_{25}$ (R-band) surface-brightness cut for J0409-56 in Figure 1c. A dotted line marks the selection ellipse, and the circled objects represent objects detected by the finder that meet color and size criteria. Two of the six circled objects lie within 2$\times$r$_{25}$, and are therefore ultimately rejected by the automated ELdot finder.

\item Finally, we examine the objects that meet the above criteria in 100 $\times$ 100 pixel (43" $\times$ 43") image cut-outs produced by the program. This step is used to reject the few remaining cosmic rays, or other spurious edge objects. 

\end{itemize}

\subsection{Sample Overview}

 Our final sample of ELdots from 93 SINGG images includes 96 compact emission-line sources with line fluxes ranging from 1.07$\times10^{-16}$ to 4.16 $\times10^{-15}$ ergs s$^{-1}$ cm$^{-2}$. 50 of the 93 SINGG systems probed contain at least one ELdot, sampling the full myriad of SINGG systems, from strongly interacting groups to quiescent spiral galaxies. However, 4 SINGG systems cover such a large angle that little or none of the image area for r/r$_{25}$ $=$ 2 -3 is covered, and so we exclude those systems from any statistical analysis.  In total, we cover a search area of 4.0 deg$^{2}$. Because the narrow band filters used by M06 vary by over a factor of 2 in bandwidth, a more relevant quantity is the volume in terms of \AA~deg$^{2}$, which we determine by multiplying the area beyond 2$\times$r/r$_{25}$ in each field by the full width at half maximum transmission of the filter as tabulated in M06. This calculation yields a total apparent surveyed volume of 165 \AA~deg$^{2}$.

 Table 1 lists all SINGG ELdots (named by the HIPASS target and ELdot number) and their positions, EWs, distance in r/r$_{25}$ from the host SINGG galaxy, \ha~ fluxes, R$-$band magnitudes and their errors, information about spectroscopic follow up (see next section) and host galaxy morphologies (see below). We find 24 ELdots in the SINGG field J0403-43, and tabulate the properties of those ELdots separately in Table 2. We performed circular aperture photometry on each ELdot using the IDL routine APER, and have accounted for photometric calibration errors (see SR1),  and sky and continuum subtraction errors in the listed flux errors. The error due to the pixel to pixel variance of the sky is the largest contributor. Eleven ELdots have no R-band emission above our detection limit of 3$\sigma$ times the continuum sky variance, and so the values in the table for the R-band magnitudes are upper limits, denoted by a ``$>$".

 Column 10 of Table 1 contains a morphological type defined as follows: Solitary (solo) signifies that this SINGG galaxy has no known companions and is free from morphological disturbances in SINGG \ha~ and continuum images, HI synthesis maps (where available), and digital sky survey images; Companion (comp) indicates a known companion or newly discovered SINGG companion for the galaxy, but no visible disturbance in galaxy structure seen in the previously mentioned set of images;  Interacting (int) are systems that are clearly disturbed in either optical or HI synthesis images or both, with or without a discernible nearby companion. The motivation for this classification scheme is twofold. First, we wanted to include information from available neutral hydrogen synthesis maps, which often show disturbance when none is visible optically. Second, we wanted to account for galaxies that have nearby companions, a relevant consideration in the context of ELdots, potential companions themselves. Since the SINGG sample was initially selected based on HI mass, there are very few early-type galaxies within it, so the traditional morphological classification scheme (spiral vs. elliptical) is largely irrelevant. The distribution of morphologies for the 89 SINGG primary galaxies searched for ELdots is as follows: 62 appear to be solitary, undisturbed systems; 17 systems are distinctly disturbed or interacting; and 10 systems have known companions but show no signs of interaction or disturbance.  For reference, the 39 systems that do not contain ELdots (and therefore are not tabulated in Table 1) have the following morphologies: 31 are solitary systems, 4 have companions, and 4 are interacting.

  \subsection{Selection Effects}
  \label{selectioneffects}
Here, we discuss some of the biases inherent in our sample selection method.  Because we define ELdots in a purely observational sense (point sources of high net emission in the SINGG narrow bandpass) we are subject to distance-related selection effects. The size selection effect, mentioned above, is such that point sources in the most distant SINGG systems could be well-resolved into dwarf galaxies were they considerably more nearby. For instance, if an ELdot in the most distant SINGG system actually has a physical size of $\sim540$ pc (the limit set by the seeing; 1.6\arcsec~ at 70 Mpc), it would be rejected in a system more nearby than $\sim30$ Mpc since its angular size would be larger than the maximum angular size in the most nearby SINGG systems (see FWHM selection criterion, above).  In practice, however, we do not see this sort of rejection. In each image there are usually between 10 and 20 objects with sizes greater than our maximum threshold, and they are invariably galaxies with no H$\alpha$ emission (this cut is applied before the NB$-$R color cut) and the SINGG galaxies themselves (sometimes detected as multiple sources with SExtractor). These objects would be subsequently rejected after implementing either the equivalent width criteria (in the case of galaxies with no H$\alpha$) or the criterion that they be outside 2$\times$r/r$_{25}$ of the SINGG galaxy (in the case of the SINGG galaxies themselves). 

It is possible that a small-sized, nearby (d $<$ 30 Mpc) SINGG  galaxy with H$\alpha$ emission could be detected as an ELdot if it were located instead in the upper distance range ( 30$-$ 70 Mpc) of the HIPASS targets. By examining all of the nearby galaxies in SR1 (62 galaxies with d $<$ 30 Mpc), we identify only two SINGG galaxies (primary HIPASS targets) that meet all the ELdot criteria except that they fall  beyond the maximum angular size threshold: ESO358- G060 (FCC302; J0345-35) and ESO444-G084 (J1337-28). These are nearby dwarf galaxies with very low R-band surface brightnesses that contain compact, luminous HII Regions. ESO358-G060 is an edge-on, LSB member of the Fornax Group \citep{drinkwater01} with a distance of $\sim20$ Mpc \citep{mould00}. The physical size of its central HII region is 850 pc, and at 70 Mpc, it would appear as a point source, with little or no underlying continuum emission. ESO444-G084 is a compact dwarf irregular galaxy that lies at a distance of 4.6 Mpc (TRGB; \nocite{karachentsev02} Karachentsev et al. 2002) and  is  only 380 kpc from M83. The physical size of the luminous HII regions in this galaxy, taken together, is approximately 600 pc, roughly the size of SINGG point sources at 70 Mpc. These two galaxies are potentially illustrative of the types of sources we find in the most distant systems.

Another distance-related caveat for the sample is that SINGG is a magnitude-limited survey. As such, the faintest HII regions in the most nearby systems would not be detected in the most distant systems. We would, however, be able to detect the ionizing flux from a single O7V star (roughly 10$^{37}$ ergs s$^{-1}$; faint end of extragalactic HII region luminosity distributions; see \nocite{oey03} Oey et al. 2003) at distances up to about 30 Mpc. At the most distant SINGG target with d$\sim70$Mpc, the limiting H$\alpha$ luminosity is roughly 10$^{38}$ ergs s$^{-1}$.  

With both the magnitude and size selection effects, 30 Mpc marks an approximate distance beyond which the sample is no longer sensitive to single-star-dominated, faint HII regions, and H$\alpha$ point sources potentially correspond to small dwarf galaxies. Outer-HIIs more distant than 30 Mpc (complexes of HII regions, dwarf galaxies) are therefore distinct  from those that lie in more nearby systems (typical galactic OB associations). We will return to this distance limit in our discussion (Section 7), where we consider a more physically homogeneous sample of outer-HIIs that, in both luminosity and size, look like typical galactic HII regions.

  \section{Spectroscopic Follow-Up}
 
ELdots, while generally having large equivalent widths in the SINGG narrow band filters, may not be H$\alpha$ emitters at the velocity of the SINGG galaxy. Other strong nebular emission lines ([OIII], H$\beta$, or [OII]) may be responsible for the comparatively high narrow-band flux if the redshifted emission line falls into the \ha~filter bandpass used to observe the target SINGG galaxy.  Therefore, we used spectroscopy to determine which ELdots are \ha~emitters at the redshift of the target galaxy and which are background sources emitting another line in the narrow bandpass. We undertook follow-up long-slit spectroscopy at multiple telescopes, sometimes as a telescope's primary program, other times as back-up observations during non-photometric conditions. We obtained follow-up spectra for 33 of the ELdots, and found background galaxies to comprise the majority of the sample. Since outer-HIIs at the redshift of the SINGG galaxy are our primary interest for this work, we chose to discontinue spectroscopic follow-up for the entire sample.\footnote{ Lessons Learned from Follow-up Spectroscopy: 
Because most ELdots are optically very faint in R-band continuum images, the object acquisition process for follow-up spectroscopy was quite difficult, and generally required blind offsets from nearby bright stars. Ultimately, we found multi-slit masks to be the most efficient way of getting spectra for these faint objects, however applying this method to the entire SINGG sample is not only costly, but a poor use of slit mask space (can hold thousands of slits in a single field) considering the presence of only 1-2 ELdots per field. Furthermore, the majority of the ELdots in SINGG are background galaxies. GALEX data is important for understanding the full context of the outer-galaxy star formation, but it also emerges as a useful tool for distinguishing between background galaxies and outer-HIIs (see Section 5.2). We recommend using follow-up spectroscopy for confirmation of local star formation only in truly ambiguous cases for which there are no GALEX data available.}  In this section we present follow-up spectroscopy for roughly half of the SR1 systems containing ELdots, and in the next section, present a more efficient photometric method for distinguishing between background galaxies and outer-HIIs. 

 Observations carried out in September 2002 and October 2003 were presented in RW04, and represent spectra taken with the Double Beam Spectrograph on the RSAA 2.3 m telescope. Subsequent runs in November 2005, May 2006, and December 2006 took place at Las Campanas Observatory on the Baade 6.5 m telescope with the Inamori Magellan Areal Camera and Spectrograph (IMACS). The long-slit spectra obtained during these runs used the 300 l/mm grating and the f/4 camera, resulting in a dispersion of 0.74 $\AA$ pixel$^{-1}$ and a spectral range of 3650 - 9740 $\AA$. Exposure times varied between 900 and 2000 seconds, depending upon the flux of the ELdot observed. Absolute flux calibration was not paramount to our goal of getting the ELdot's velocity.  
\begin{figure*}[p!]
\begin{center}
\vspace{-0.01in}
\includegraphics[height=2.0in]{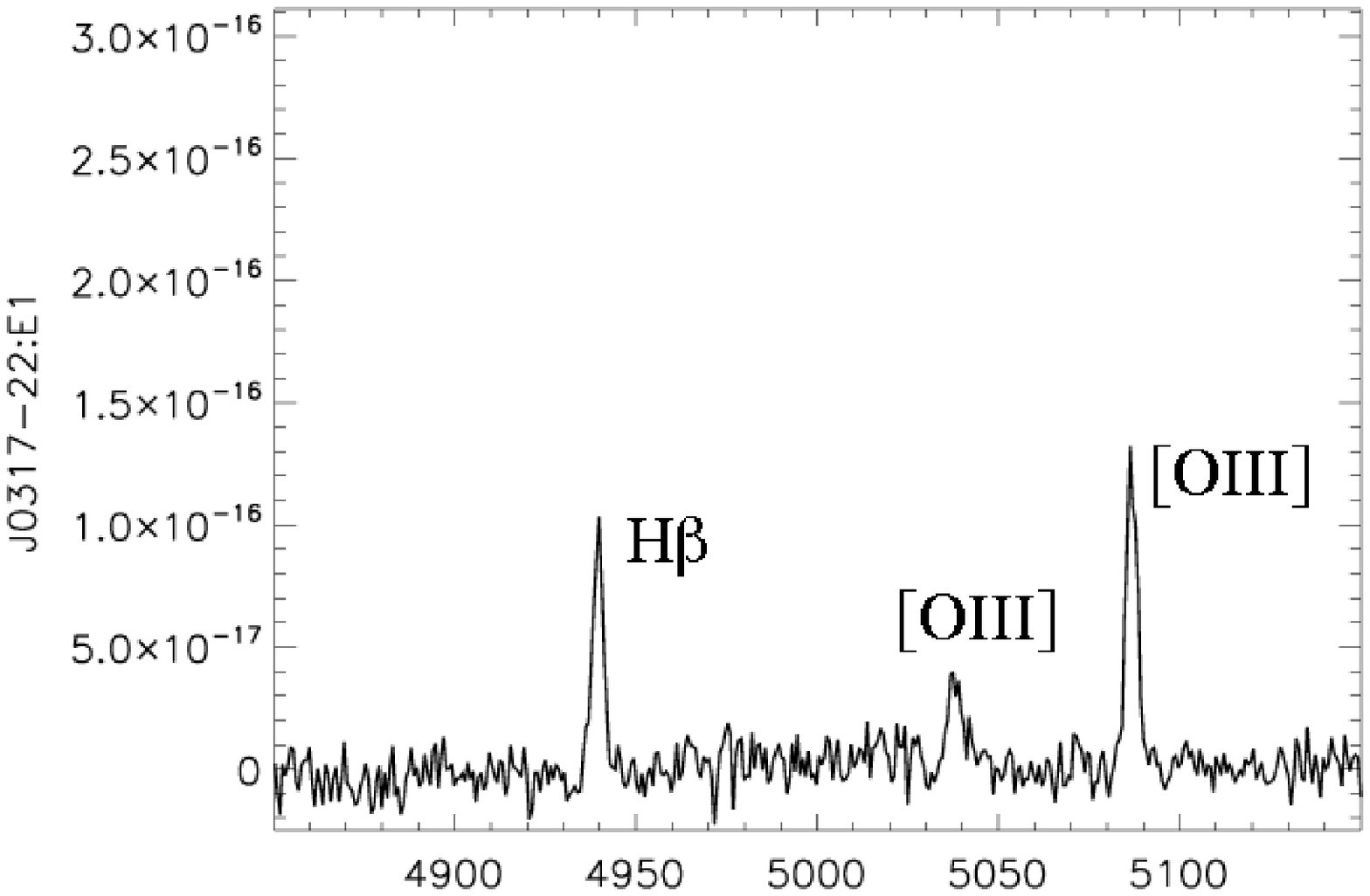}
\includegraphics[height=2.0in]{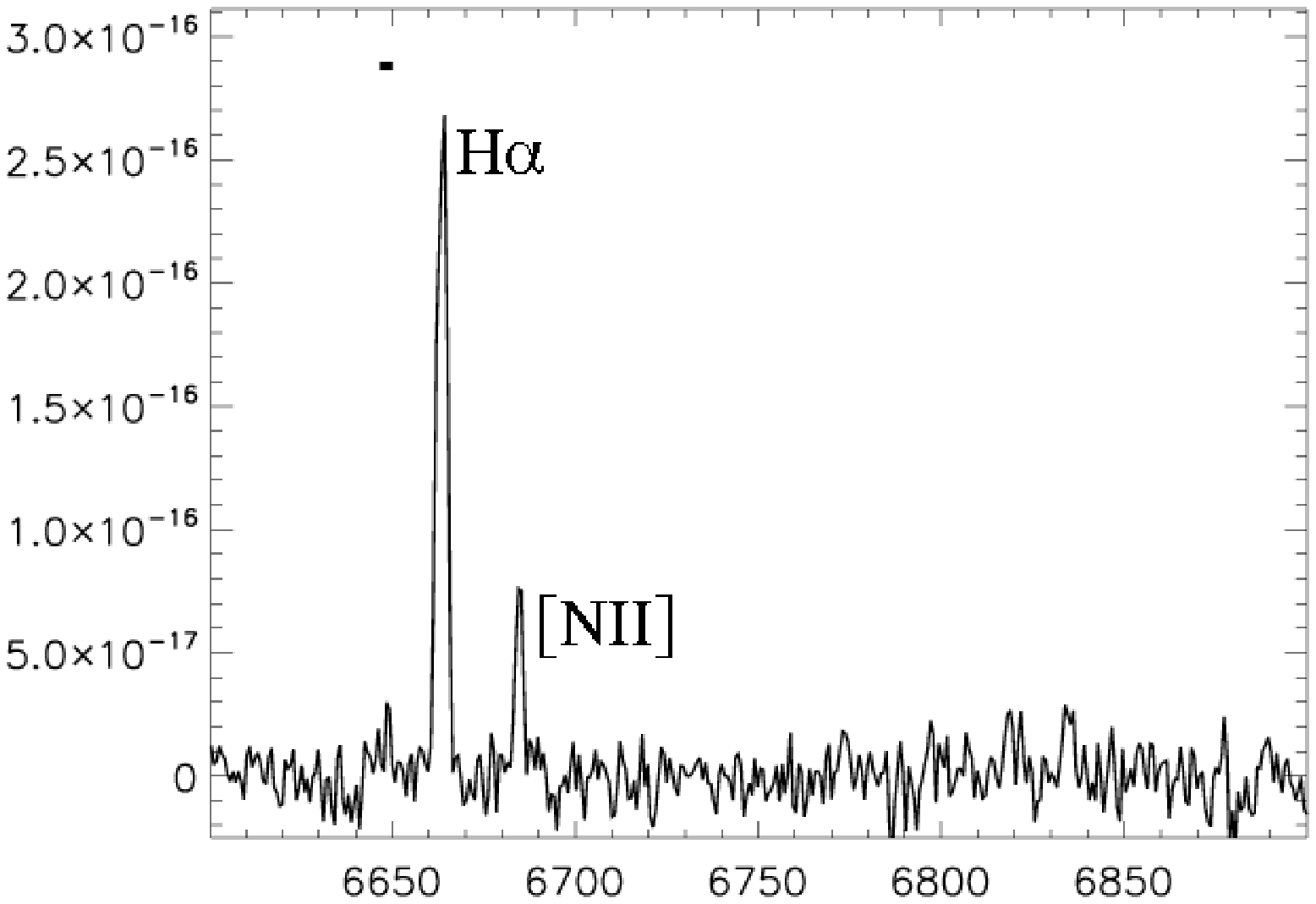}
\includegraphics[height=2.0in]{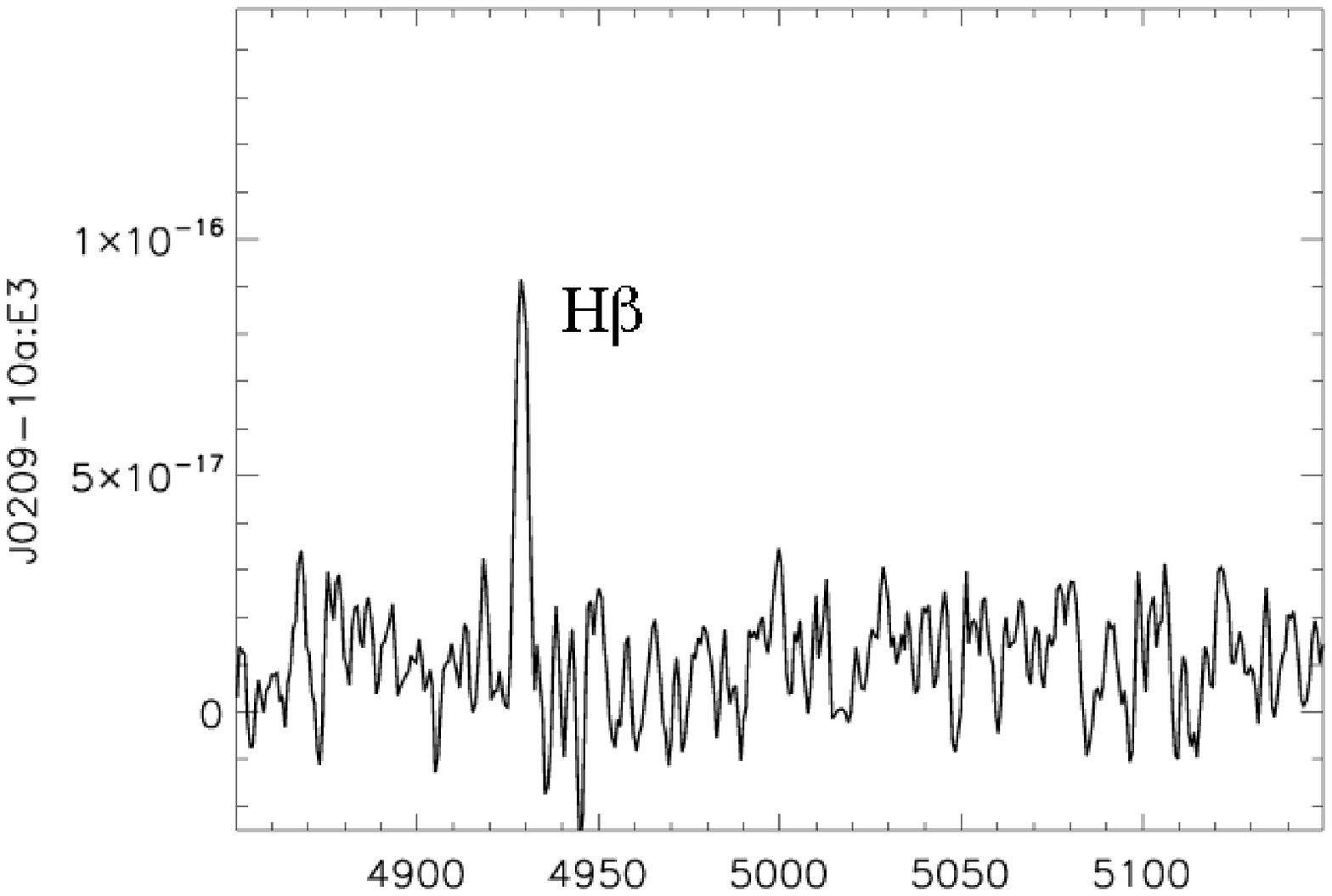}
\includegraphics[height=2.0in]{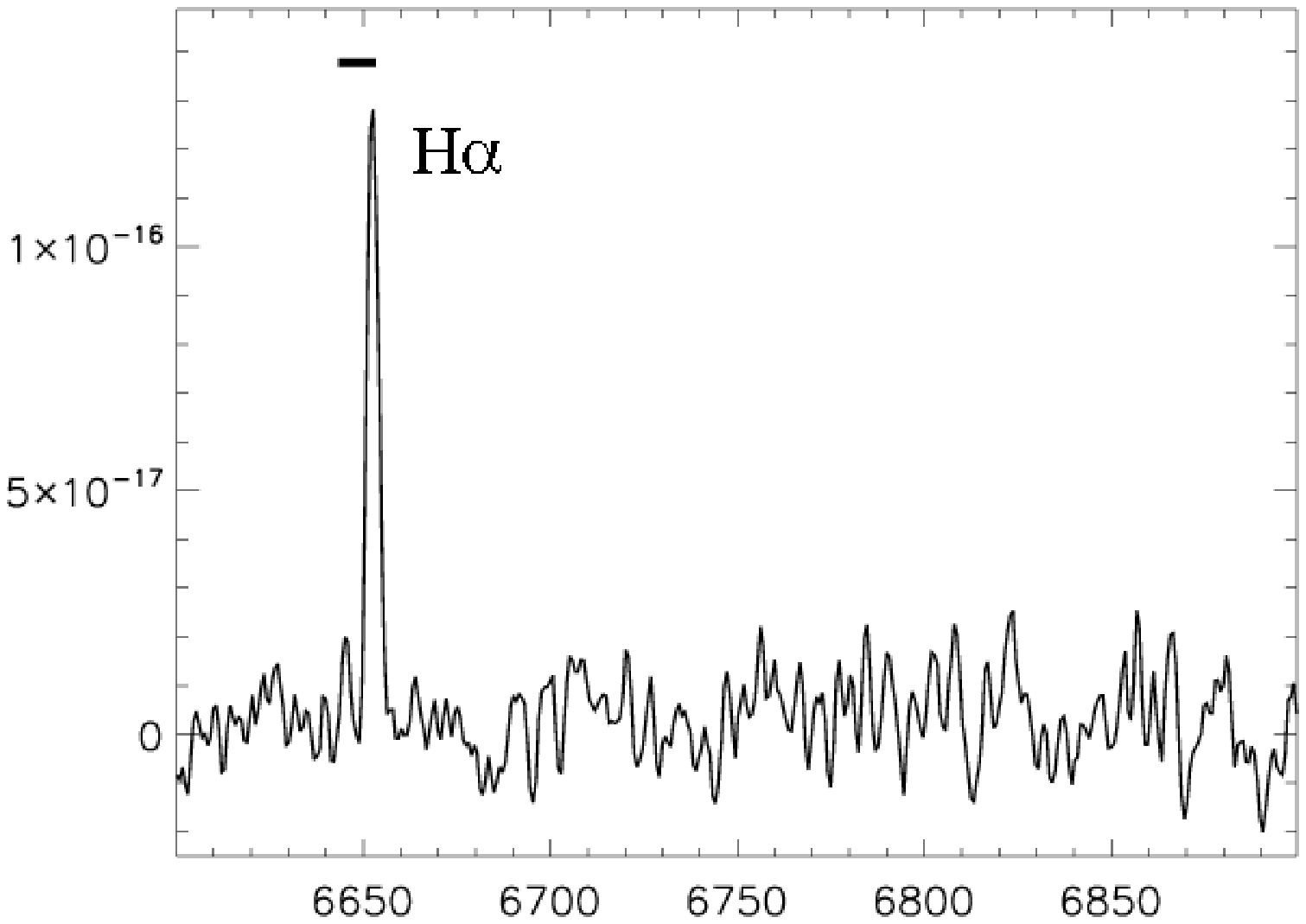}
\includegraphics[height=2.0in]{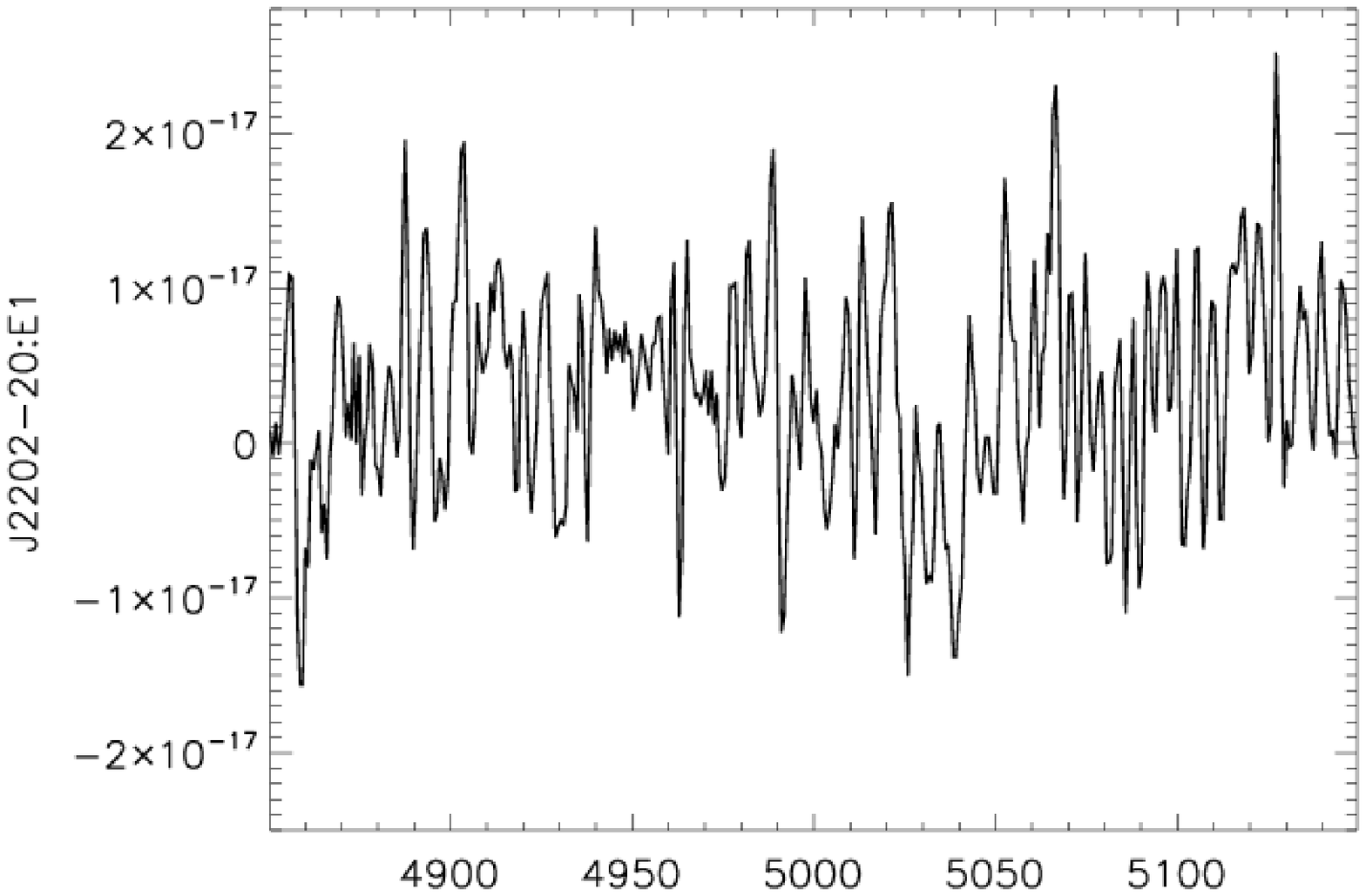}
\includegraphics[height=2.0in]{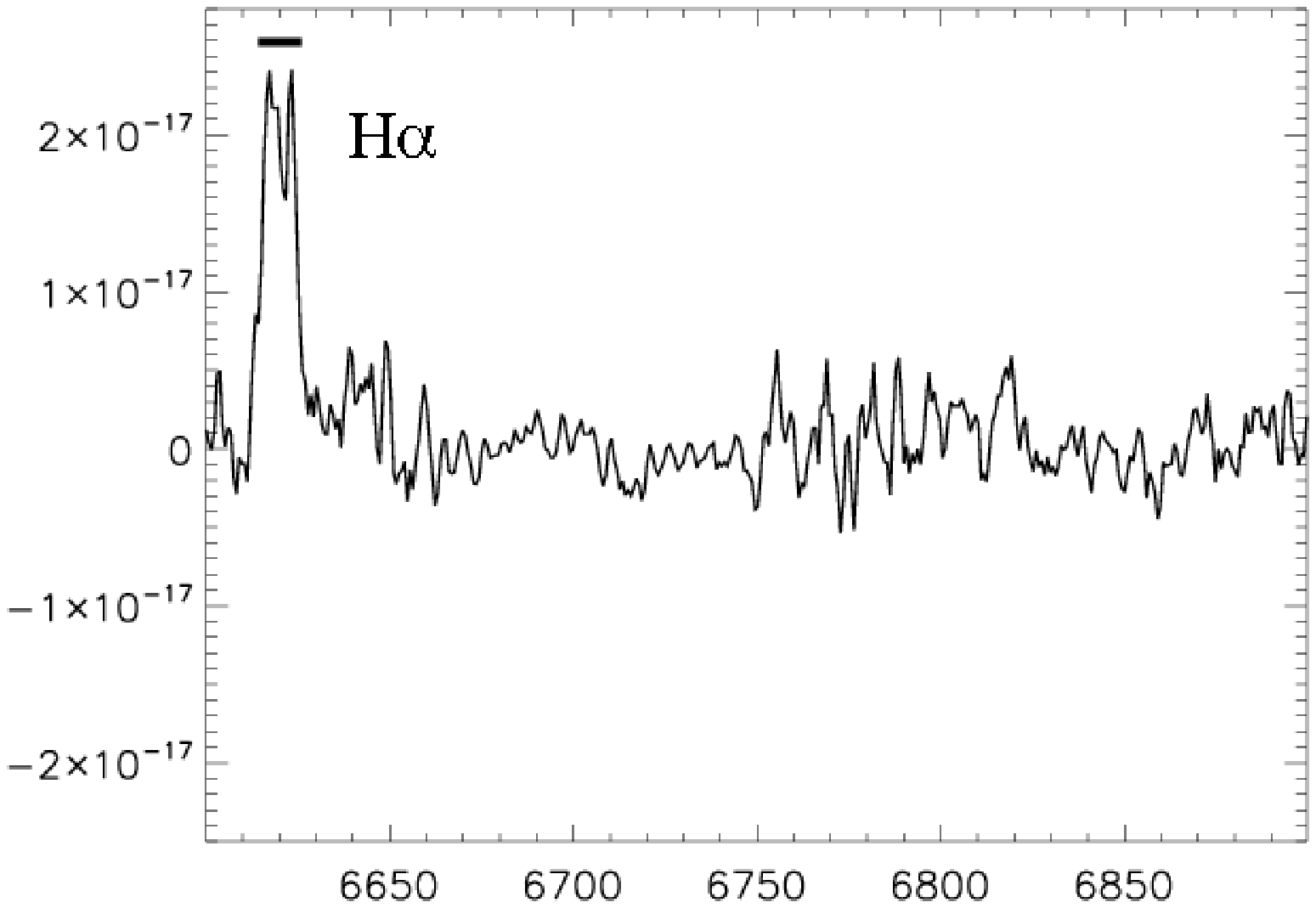}
\end{center}
\caption{\label{spectra}Follow-up spectra of three ELdots classified as outer-HIIs. Bold horizontal lines span the FWHM of the 21cm line profiles for each HIPASS target, and are placed at the expected velocity of the \ha~emission line.  Top: emission-line spectrum of J0317-22:E1 showing [OIII] and H$\beta$ emission lines in the blue, and \ha~and [NII] in the red. The \ha~emission of the ELdot is offset from the potential host galaxy by about  20\AA, or 700 km/s. Middle: Spectrum of J0209-10a:E3 showing two emission lines, best fit by H$\beta$  and \ha. Bottom: Single-line spectrum of J2202-20:E1 showing coincidence of the main emission line with the HI profile.}
\end{figure*}
 We spectroscopically confirm eight ELdots to be \ha~ emitters at comparable velocities to their apparent host galaxies. Five of these outer-HIIs were presented in RW04. Figure \ref{spectra} shows the reduced and calibrated spectra for three newly confirmed outer-HIIs: J0317-22:E1, J0209-10a:E3, and J2202-20:E1, with the results discussed below.  For reference, the bold, horizontal lines span the FWHM of the observed HIPASS HI profiles for the SINGG host galaxies, and are placed in wavelength space at the expected velocity of the \ha~emission line. \begin{itemize}
 \item{In the spectrum of  J0317-22:E1, we have detected H$\beta$, [OIII] $\lambda$ 4959 \AA~and 5007\AA, H$\alpha$ and [NII] $\lambda$ 6548\AA~ and $\lambda$ 6583 \AA. The recession velocity measured from the H$\alpha$ emission line is $\sim4700$ km/s ($\lambda=$ 6664 $\AA$), quite a bit offset from the HI velocity of the host galaxy (ESO 481-G017) at $\sim3920$ km/s (spanning $\sim3850$ km/s to $\sim4000$ km/s).  HI synthesis maps recently obtained with the VLA show a small HI cloud at the velocity of the ELdot, but no direct gaseous connection to the main galaxy \citep{nitza09}. We show the SINGG image of this system, along with the location of two ELdots in the field in Figure \ref{j0317-22_ELdots}. We obtained follow-up spectroscopy for J0317-22:E2 as well, and multiple lines confirmed it is a background emitter at z$\sim0.3$. }
 \item{The spectrum of J0209-10a:E3 reveals only two emission lines, H$\beta$ and H$\alpha$, and is the second outer-HII in the J0209-10a system (HCG 16). The recession velocity from the H$\alpha$ emission line is  4160 km/s ($\lambda=$ 6653 $\AA$), comparable to that of the SINGG host galaxies, measured from the HI profile, at $\sim3900$ km/s. We show the SINGG image of this system, with the two spectroscopically-confirmed outer-HIIs circled and labeled in Figure \ref{hcg16_ELdots}. As it lies south of  the image area shown in Figure \ref{hcg16_ELdots}, we do not show the location of J0209-10a:E1, for which we did not obtain a spectrum.}
 \item{ Finally, we detect only one emission line for J2202-20:E1, but consider it very likely to be H$\alpha$. The recession velocity measured from the emission line at 6620 $\AA$, 2660 km/s, matches the HI velocity of the host galaxy (NGC 7184), 2620 km/s, very well. Furthermore, the ELdot is located on the major axis of the optical disk, on the northeasterly receding edge \citep{vanderkruit}, and could easily be an isolated extension of a spiral arm (see Figure \ref{j2202-20_eldots}). In addition to J2202-20:E1, we show the location of J2202-20:E2 in  Figure \ref{j2202-20_eldots}. We did not obtain a follow-up spectrum for this ELdot. }
 \end{itemize}
 \begin{figure*}
\epsscale{1.0}
\plotone{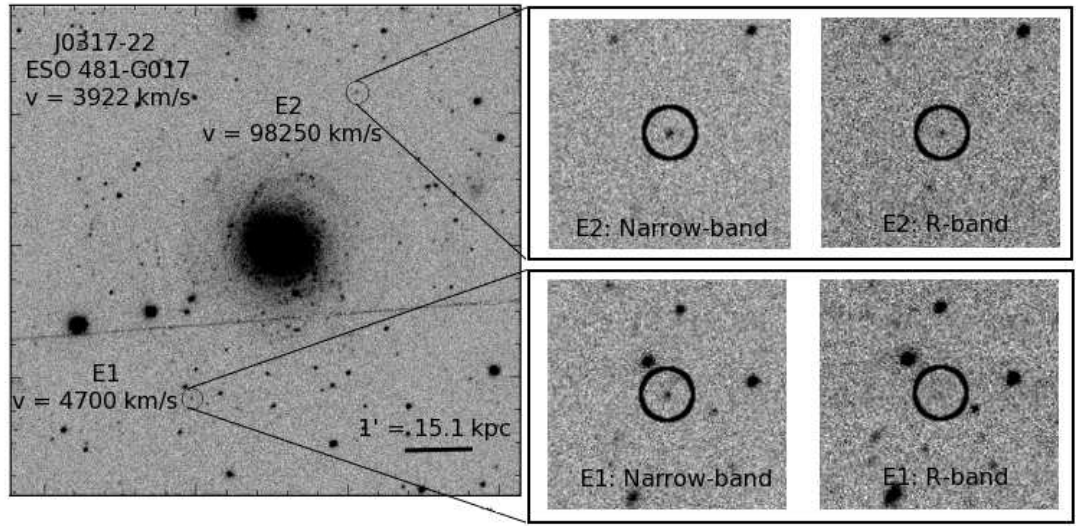}
\figcaption{\label{j0317-22_ELdots} SINGG \ha~ image of the field J0317-22 (ESO 481-G017), with its two ELdots circled and labeled. On the right, close-up views of the ELdots in both narrow-band and R-band filters. Follow-up spectra have confirmed that E1 is an outer-HII, at roughly the same distance as the SINGG galaxy, while E2 is an [OIII] emitter at a redshift of 0.3. }
\end{figure*}

\begin{figure*}
\epsscale{1.0}
\plotone{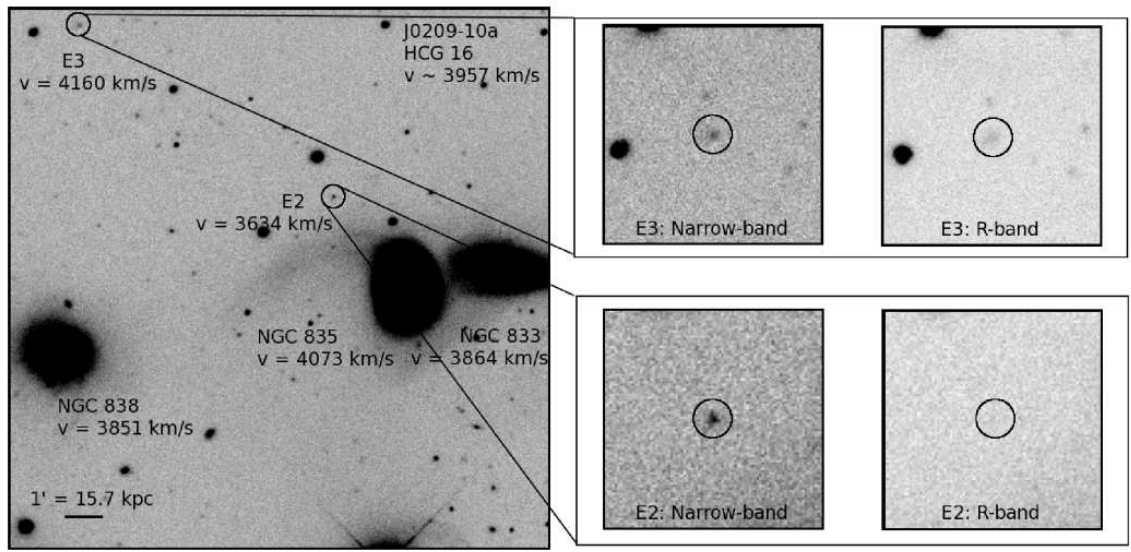}
\figcaption{\label{hcg16_ELdots} SINGG \ha~ image of the field J0209-10a (HCG 16), with its two ELdots circled and labeled. On the right, close-up views of the ELdots in both narrow-band and R-band filters. J0209-10a:E1 is not shown in the image, as it is considerably further south (off the image area shown) than E2 and E3. Follow-up spectra have confirmed that both E2 and E3 are outer-HIIs, at roughly the same distance as the compact group. }
\end{figure*}

\begin{figure*}
\epsscale{1.0}
\plotone{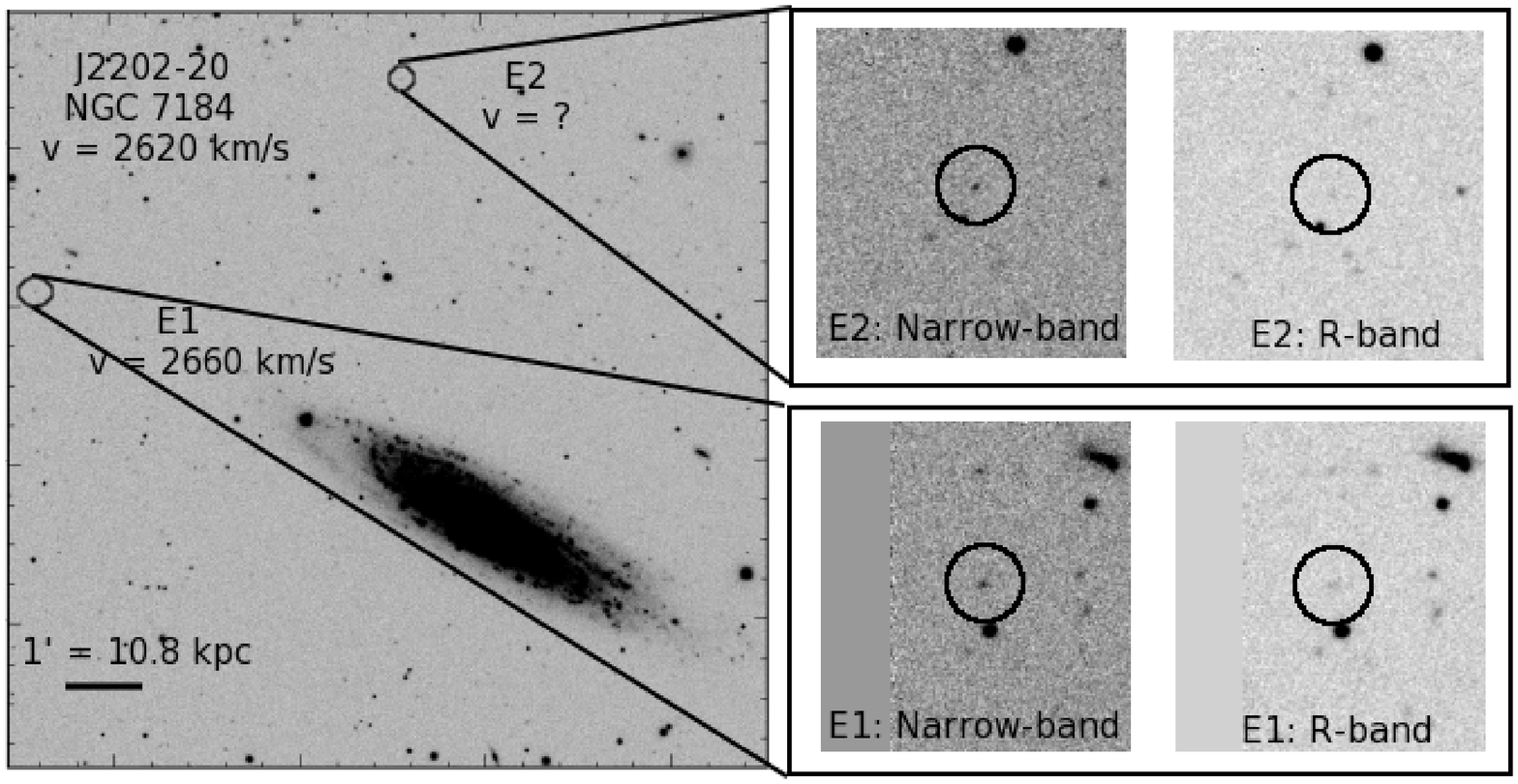}
\figcaption{\label{j2202-20_eldots} SINGG \ha~ image of the field J2202-20 (NGC 7184), with its two ELdots circled and labeled. Same as Figure \ref{j0317-22_ELdots}, we show close-up views of the ELdots in both narrow-band and R-band filters on the right. In this case, a follow-up spectrum of  E1, combined with HI data of \cite{vanderkruit} strongly indicates that E1 is an outer-HII, at roughly the same distance as the SINGG galaxy. We do not have spectroscopic follow-up data for E2. }
\end{figure*}

\begin{figure}
\epsscale{0.70}
\plotone{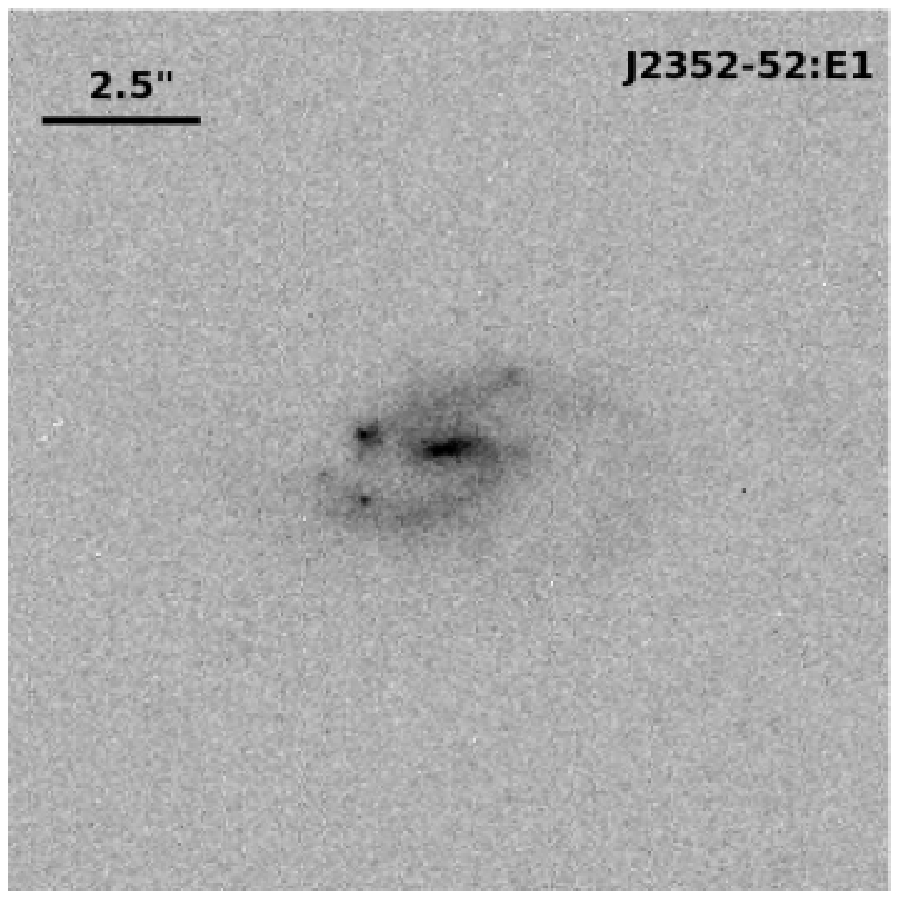}
\figcaption{\label{eso149} HST HRC image of the ELdot in the HIPASS target J2352-52. This image, in the F814W bandpass, reveals that J2352-52:E1 is a background galaxy with spiral structure. RW04 detected only one emission line in this object's long-slit sprectrum, and therefore could not confirm its nature. Considering the angular size of the galaxy (2.5"), the emission line could be either H$\beta$ at z$\sim0.4$ or [OIII] at z$\sim0.3$, making its diameter either 16 kpc or 20 kpc. }
\end{figure}

 Two other ELdots were presented by RW04 for which we now have additional information. We detected multiple emission lines of the ELdot J0409-56:E3 using an IMACS multi-slit mask, confirming its association with SINGG galaxy NGC 1533, but will present its spectrum, along with others, in a forthcoming study on outer-HII region metallicities. The ELdot J2352-52:E1, initially presented as a single-line detection in RW04, appears to be a red background galaxy in HST High Resolution Channel images. These {\it{HST}} ACS/HRC observations were carried out in October 2004 in three filters: F250W (5910 s); F555W (2796 s); and F814W (2908 s). For more information on data reduction, see \cite{werk08}. We show the F814W image in Figure \ref{eso149} that reveals J2352-52:E1 to be a background galaxy with spiral structure and ongoing star formation.  We do not present the individual spectra of the 23 spectroscopically-confirmed background emitters and one foreground M star in this paper.  20 of these background emitters are emission line galaxies at z$\sim0.3$, 2 are H$\beta$-emitters at z$\sim0.4$, and 1 is an [OII]-emitter at z$\sim0.8$.  
 
\section{Comparison with deep GALEX data}
\label{galex}

While \ha~emission typically  signifies the presence of  O stars with M$_{\star}\geq20$M$_{\odot}$, UV emission traces stellar light from both O and B stars with M$_{\star}\geq3$M$_{\odot}$. Therefore, deep GALEX images of the outer-HIIs can help provide a more detailed picture of the star formation occurring at large galactocentric distances.  In turn, the H$\alpha$ images of the XUV disks provide an important constraint on the youngest, most massive stellar populations at large radii. Recent results have shown that extended UV  (XUV) emission (beyond optical) is relatively common in gas-rich nearby galaxies imaged with GALEX \citep{Thilker07}. This star formation at large galactocentric radii and how it compares to that within the main optical body is a key part in assembling a complete picture of  the star formation process on galactic scales.

We searched the Multimission Archive at Space Telescope Science Institute (MAST) for GALEX images of ELdots with exposure times longer than $\sim1000$ seconds.  We did this with two primary goals in mind: first, UV images of  outer-HIIs broaden our understanding of their environment and outer-disk star formation in general; second, UV images of  ELdots may reveal a method for distinguishing outer-HIIs from background galaxies that is less observationally expensive than follow-up spectroscopy. 16 of the SINGG fields containing ELdots have long-exposure GALEX data available from the MAST archive. Most were obtained through various  GALEX Guest Investigator (GI) programs, while a few were part of the Nearby Galaxy Survey (NGS) or the Medium Imaging Survey (MIS). Some of the galaxies are also part of the Survey of Ultraviolet emission in Neutral Gas Galaxies (SUNGG), a sub-sample of SINGG targets imaged by GALEX as part of a Cycle 1 Legacy Survey.  14  of the 16 SINGG fields, along with GALEX program IDs and exposure times are given in Table 3.  The two SINGG fields we do not include in this table are J0459-26 and J0403-43. J0459-26 is excluded because the background  ELdot in this field is detected in neither the FUV nor the NUV GALEX images, although it does have long-exposure GI and MIS GALEX data available.  J0403-43 is excluded because the large number of ELdots in this system justifies a separate table (Table 2). We do not detect J0005-28:E1 or J0403-43:E19 in their FUV deep GALEX images but include the magnitudes for these objects as upper limits based on the background of the GALEX image (3$\sigma$ detection thresholds). The FUV bandpass of GALEX has an effective wavelength of 1516 \AA, and the NUV GALEX bandpass has an effective wavelength of 2267 \AA.

\subsection{GALEX Morphology and Outlying HII Regions}

A number of studies that examine the relation between UV and H$\alpha$ emission in the outer parts of galaxies claim a deficiency of massive, young O stars (e.g. Gil de Paz et al. 2005; Meurer et al. 2009 \nocite{gildepaz05, meurer09}). This deficiency is still under debate, and could be due to stochastic effects in a low star formation efficiency environment, a different stellar or cluster initial mass function at larger radii, or could also be a product of the lack of a uniform, high-quality H$\alpha$ dataset. Whatever the case, it is important to examine the connection between outer-HIIs and XUV emission. As we have seen, there are certainly HII regions present in the outer parts of galaxies, and whether or not they are part of a larger population of extended star formation is quite relevant to answering some of the questions raised in recent studies regarding their paucity. Here, we examine the UV morphologies of four SINGG systems with outer-HIIs. Compared to the SINGG images, the GALEX images have a psf that is 3 times as broad (4.5$\arcsec$ versus 1.5$\arcsec$), so their quality appears sightly degraded. Nonetheless, in these four cases, we see evidence that outer-HIIs are indeed part of a more extended population of UV-bright clusters. 
 
In this section, we refer to the XUV morphological types presented by \cite{Thilker07}. XUV disks display two typical morphologies that \cite{Thilker07}  classifies as either Type 1 or Type 2. Type 1 objects are classified as having clustered UV-bright/optically faint sources in their outer parts, while Type 2 objects have blue FUV$ - $NIR colors and XUV emission organized in a large disk structure. Type 1 XUV galaxies are often found in interacting systems, and a truncation radius for the star formation is not well-defined. Conversely, Type 2 disks have an obvious end to their outer-disk star formation, which is happening in a LSB disk lacking a significant older stellar population. It is possible for a galaxy to be a mixed-type as well, in that it matches both sets of criteria.

\begin{figure*}
\epsscale{0.75}
\plotone{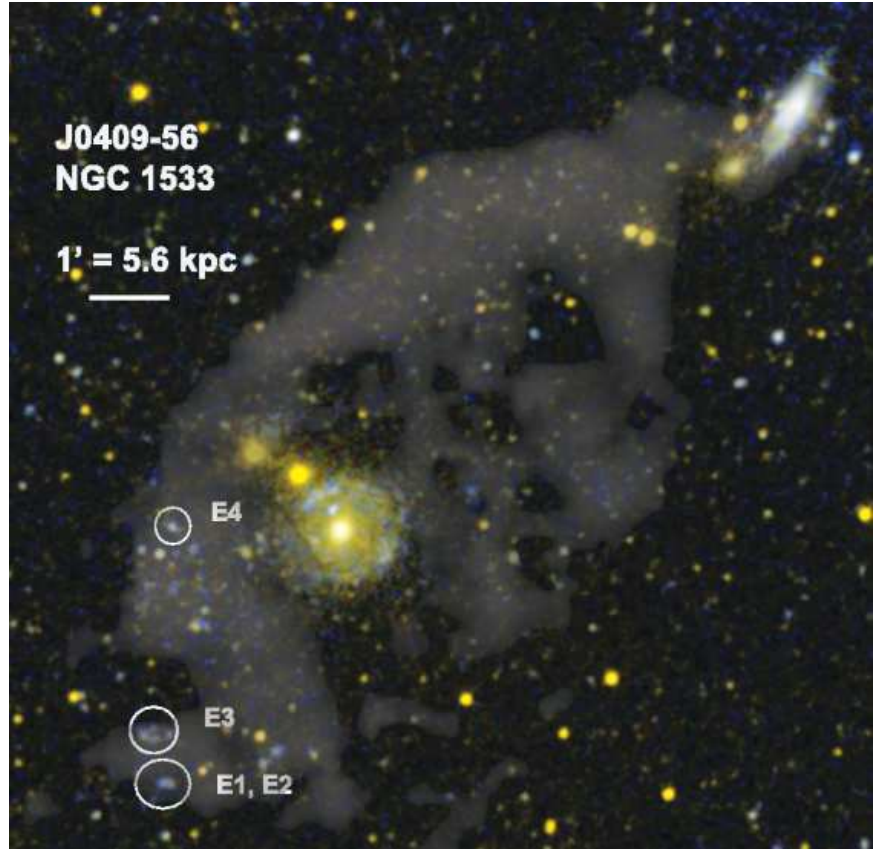}
\figcaption{\label{ngc1533galex} GALEX two-color FUV and NUV image of the HIPASS target J0409-56, with its four ELdots circled and labeled. FUV is in blue and NUV is in yellow. The clouded area shown marks the location of  an HI column density of at least 1.5 $\times$ 10$^{20}$ cm$^{-2}$ from an ATCA 21-cm synthesis map. We show the physical scale in the top left corner of the image.  }
\end{figure*}

We show the color GALEX image of  NGC1533 (HIPASS target J0409-56) that contains 4 outer-HIIs in Figure \ref{ngc1533galex}. The visible gray cloudiness surrounding NGC 1533 in Figure 7 traces the HI distribution at roughly 1.5$\times10^{20}$ cm$^{-2}$, and is based on ATCA HI contours previously presented by \cite{emmaconf}.  The SE portion of the HI cloud appears to contain a population of blue, clustered sources.  While the \ha~ emission in this area is limited to the outer-HIIs, the UV emission is more prevalent, and indicates ongoing star formation outside UV and optical threshold contours. NGC 1533 is a Type 1 XUV galaxy, with bright UV-emission complexes outside the anticipated location of the star formation threshold.  The optical extent (r$_{25}$) of NGC 1533 is only 10 kpc, based on the distance to NGC 1533 (19.4 Mpc; \nocite{deGraaf07} DeGraaff et al. 2007). The XUV emission extends to greater than 30 kpc, filling the extended HI structures surrounding NGC 1533. 

\begin{figure*}
\epsscale{1.0}
\plotone{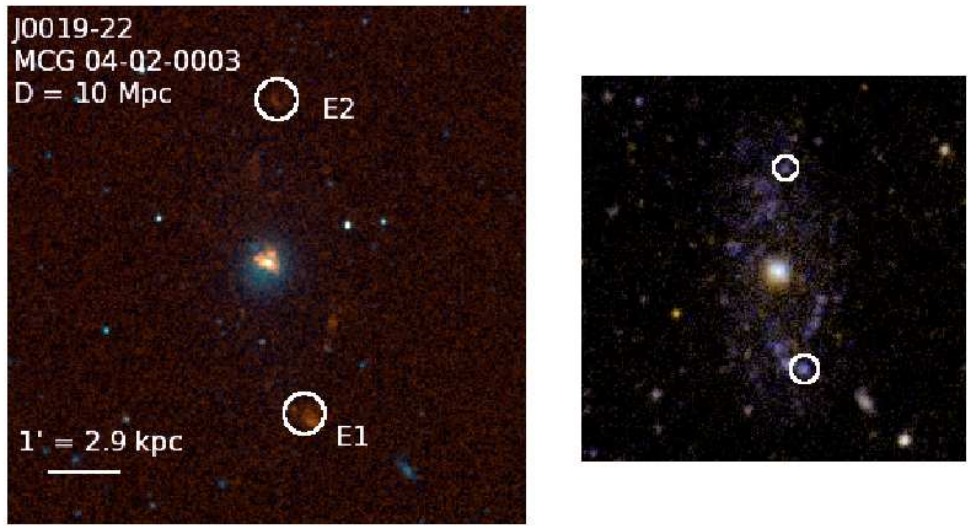}
\figcaption{\label{hafuvj0019} Left: SINGG color image (H$\alpha$ = red, R-band = blue) of the field J0019-22 containing the galaxy MCG-04-02-0003, with its two ELdots circled and labeled. Right: GALEX two-color FUV (blue) and NUV (yellow) image of the same galaxy. The ELdots are associated with two of the many  clusters bright in the FUV that lie in an extended disk around the main galaxy.}
\end{figure*}

\begin{figure*}
\epsscale{1.0}
\plotone{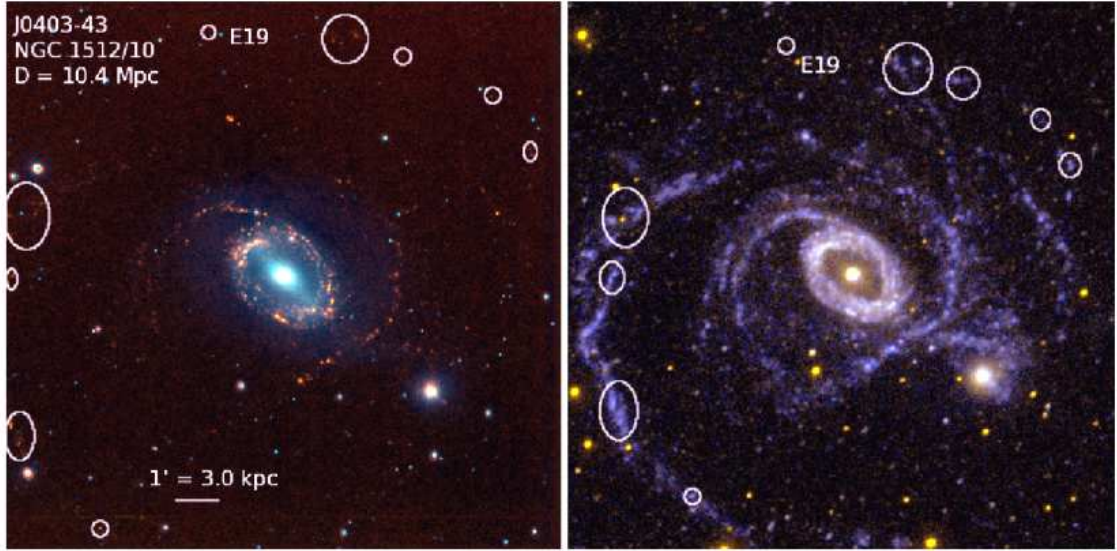}
\figcaption{\label{ngc1512} Left : SINGG color image of J0403-43 (NGC 1512/10 system), with \ha~ in red and R-band in blue. The 24 ELdots of this system are circled. Right: GALEX  color image of the same system with FUV in blue and NUV in yellow.  As noted in the text and Table 2, the lower resolution of the GALEX images sometimes results in multiple SINGG ELdots detected as a single UV-complex. We see this effect for many of the ELdots in this field. We have labeled J0403-43:E19 in both images to show that there is no FUV emission associated with this ELdot, a likely background galaxy. }
\end{figure*}

\begin{figure*}
\epsscale{1.0}
\plotone{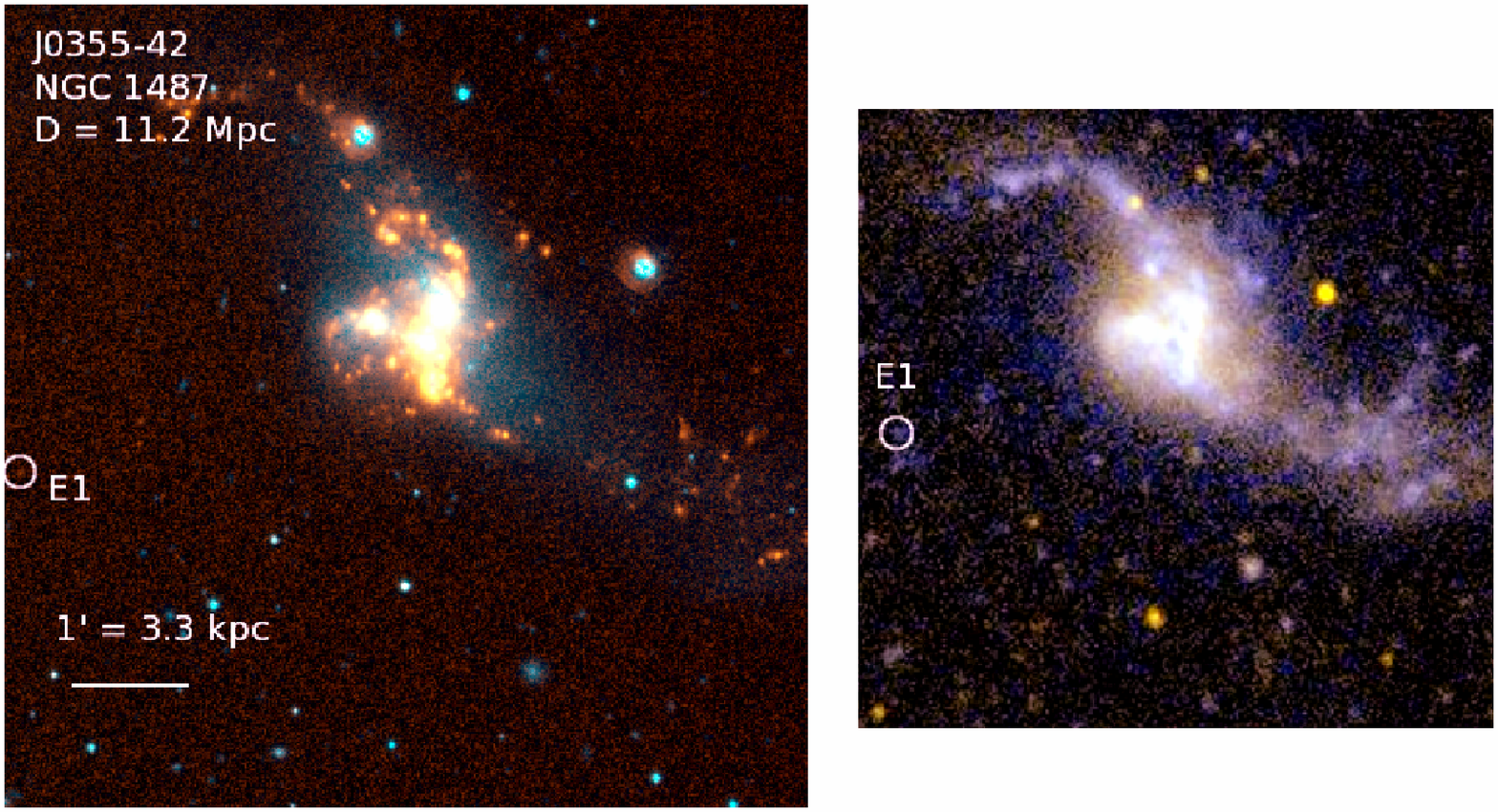}
\figcaption{\label{ngc1487} Right: SINGG color image of J0355-42 (NGC 1487), with \ha~ in red and R-band in blue.   Left: GALEX color image of the same field of view as the SINGG image on the left. On the GALEX panel, blue shows FUV while yellow shows NUV. The ELdot in this field is circled on the left-hand side of each image.}
\end{figure*}

Three other systems with ELdots exhibit Type 1 XUV emission: MCG-04-02-0003 (HIPASS target J0019-22), the NGC 1512/10 interacting pair (HIPASS target J0403-43), and NGC 1487 (HIPASS target J0355-42). We present both SINGG (H$\alpha$ and R-band) and GALEX (NUV and FUV) color images of these systems in Figures \ref{hafuvj0019} - \ref{ngc1487} for comparisons of the H$\alpha$ and UV emission.  In MCG-04-02-0003 (see Figure \ref{hafuvj0019}, the two ELdots lie at the edges of a Type 1 XUV disk that is completely invisible in the R-band, just barely discernable in \ha, and very bright in the FUV. Combined with their blue FUV$-$NUV colors (see next section), the environment of the ELdots in this system indicates that they are indeed outer-HIIs associated with MCG-04-02-0003.   In Figure \ref{ngc1512} (the NGC 1512/10 system; left panel), the ELdots stand out as H$\alpha$ (red) point sources arranged in an extended spiral structure. On the right-hand panel of Figure \ref{ngc1512}, we see that NGC 1512 has a large Type 1 XUV disk (r $>$ 30 kpc), that extends beyond the edges of the SINGG image. Its properties are presented fully in \cite{Thilker07} and also discussed extensively by Koribalski \& Lopez-Sanchez (2009). We conclude that the vast majority of ELdots in this system are outer-HIIs, given both their abundance in the SINGG image and their location in a large and bright Type 1 XUV disk. And finally, in Figure \ref{ngc1487}, the single ELdot in this system lies in a larger stellar complex with a blue FUV$-$NUV color that is part of the Type 1 XUV emission associated with NGC 1487. Again, given its blue color and surroundings of FUV-bright stellar complexes, we consider this ELdot to be an outer-HII. 

\subsection{UV and Optical Photometry}

Using deep GALEX archival data, we examine potential differences in UV flux and color between background galaxy ELdots and outer-HIIs. Through position-matching, we find the ELdots in the GALEX images and source catalogs, and list their Galactic foreground reddening-corrected \citep{schlegel98} FUV magnitudes and FUV-NUV colors in Tables 2 and 3. The photometric measurements come from combined, flux calibrated, background-subtracted images taken from the GALEX pipeline \citep{Morrissey07}.  We have separated the ELdots in Table 3 by the status of their spectroscopic confirmation. In Table 2, we have placed horizontal lines delineating the multiple ELdots detected as single UV-complexes in the GALEX image due to the lower resolution of the GALEX images compared to the SINGG images (4.5$\arcsec$ versus 1.5$\arcsec$; see Figure \ref{ngc1512}). 
 
The median FUV$-$NUV color of the seven background galaxies is 0.20, while three spectroscopically confirmed outer-HIIs have bluer UV colors, from 0.03 to -0.79. When we include the large number of outer-HIIs in J0403-43,  the two in J0019-22, and the one in J0355-42, the color separation is more pronounced.  All of the outer-HIIs in J0403-43 have FUV$-$NUV colors well below zero, except for J0403-43:E19, which has fairly red FUV$-$NUV color of 0.56.  The two outer-HIIs in J0019-22 have moderately blue colors of 0.18 and 0.10, while the outer-HII in J0355-42 has a color of -0.32. 

Although the outer-HIIs have blue FUV$-$NUV colors, a blue FUV$-$NUV color does not necessarily indicate an ELdot is an outer HII. Three of the seven background galaxy ELdots, J0512-32:E1, J0506-31:E1 and J2149-60:E1 have quite blue colors (FUV$-$NUV = 0.05, -0.30, and -0.55, respectively). On the other hand,  a red FUV$-$NUV color ($>$ 0.25) appears to positively indicate a background source. The reddest outer-HII has a FUV$-$NUV color of 0.18, and the vast majority ($\sim90$\%) of outer-HIIs have FUV$-$NUV colors $<0$. Of the ten ELdots with UV photometry but without spectra, six have red colors consistent with  background sources. Furthermore, the red ELdot, J0403-43:E19, may be a background galaxy in a field replete with outer HIIs, especially since it is not located in one of the well-defined spiral arms of NGC1512/10 (see Figure \ref{ngc1512}) and its narrow-band line emission and equivalent width are the lowest of the J0403-43 ELdots (see next section).  We do not include this ELdot in our sample of outer-HIIs. 

\begin{figure}
\centering
\epsscale{1.25}
\plotone{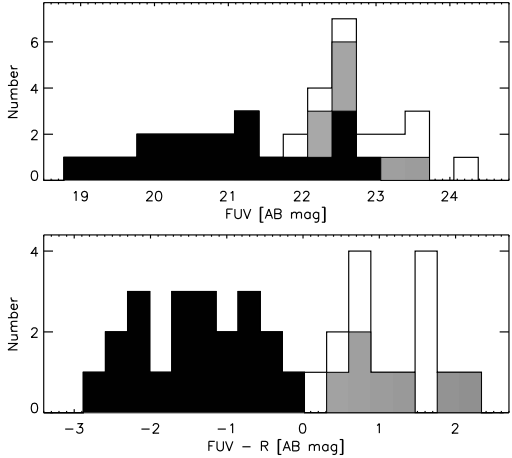}
\figcaption{\label{galexcolors} Two histograms showing the SINGG and GALEX photometric properties of all SINGG ELdots having both datasets available. In both panels, the black represents outer-HIIs, the gray represents background galaxies, and the clear area represents the ELdots that have not yet been confirmed to be either. Top: The distribution of GALEX FUV magnitudes. Bottom:  The distribution of GALEX FUV$-$R colors. }
\end{figure}

 The two histograms in Figure \ref{galexcolors} show the distribution of GALEX FUV magnitudes (top panel) and FUV$-$R colors (bottom panel) of the ELdots in the SINGG fields for which deep GALEX images are available. The  outer-HIIs (black histogram) have the brightest FUV magnitudes and the bluest FUV$-$R colors, while the background galaxies and ELdots without spectroscopic follow-up (gray and clear histograms, respectively) are faintest in FUV and reddest in FUV$-$R. The distribution of FUV$-$R is distinctly bimodal for the two primary types of ELdots. The outer-HIIs stand out as having very blue FUV$-$R colors, due to their faint continua and bright FUV emission. 

\begin{figure}
\centering
\epsscale{1.25}
\plotone{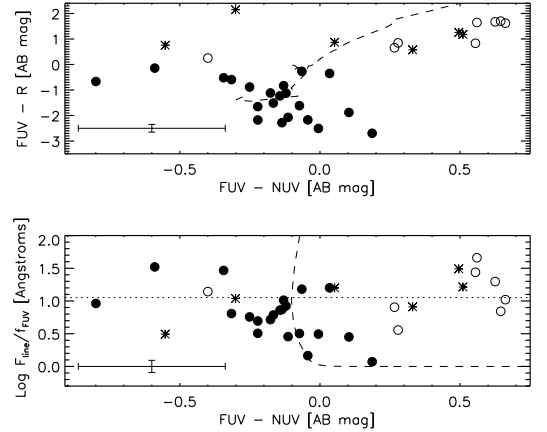}
\figcaption{\label{galexcolorcolor} SINGG and GALEX color-color plots of the ELdots having both datasets available. Filled circles mark outer-HIIs, open circles mark ELdots without follow-up, and the asterisks show the background galaxies. The dashed lines show the tracks for a single-burst stellar population with M$_{up}$= 100 M$_{\odot}$ and solar metallicity generated by Starburst99 for reference. On both plots, we show the mean error for each quantity in the lower left-hand corner.  Top: FUV$ -$NUV versus FUV$ -$ R in AB magnitudes  Bottom:  FUV$-$NUV versus SINGG line flux divided by the GALEX FUV flux density, in units of \AA.  The dotted, straight line represents a fiducial value of F$_{H\alpha}$/f$_{FUV}$, for which stars form at a constant rate with M$_{up}$ = 100 in a Salpeter IMF.}
\end{figure}

Figure \ref{galexcolorcolor} shows two plots that display the separation of background galaxies and outer-HIIs in FUV and optical color-color space. On the top panel, we plot the FUV$-$NUV color versus the FUV$-$R color, and on the bottom panel we show the FUV$-$NUV color versus the ratio of the SINGG narrow-band flux (F$_{line}$ in units of ergs s$^{-1}$ cm$^{-2}$) divided by the GALEX FUV flux density (f$_{FUV}$ in units of  ergs s$^{-1}$ cm$^{-2}$ $\AA^{-1}$) that has units of \AA. The dashed lines on both panels show the color evolution of a single-burst stellar population with a Salpeter IMF (M$_{up}=100$ M$_{\odot}$) and solar metallicity.  We note that this  model cannot accurately represent the stellar populations of the outer-HIIs, as their total masses range between 200 - 2000 M$_{\odot}$, significantly less stellar mass than is required to fully populate the stellar IMF \citep{werk08}. At these masses, stochastic effects dominate the results, and the models become degenerate. We show the Starburst99 tracks simply for reference in comparison with the numerous other works that do so. 

Most of the  background galaxies (asterisks) have redder FUV$ -$NUV colors than outer-HIIs, in addition to their redder FUV$ -$R colors, seen in the top panel of Figure \ref{galexcolorcolor}. On the bottom panel, outer-HIIs have both lower F$_{line}$/f$_{FUV}$, on average, and bluer FUV $-$ NUV colors. Both of these figures show that, given the distinction between outer-HIIs and background galaxies in color space, most of the remaining ELdots in our sample from the bottom half of Table 3 are  likely to be background ELdots. 

For reference, the dotted line on the bottom panel of Figure \ref{galexcolorcolor} marks the expected value of F$_{H\alpha}$/f$_{FUV}$ for a stellar population continuously forming stars that obeys a Salpeter IMF with M$_{up}$ = 100 M$_{\odot}$. All but four of the outer-HIIs lie below this line. A recent study by \cite{meurer09} challenges the notion of a universal IMF using integrated \ha~and UV emission  measurements for a sample of $\sim100$ SINGG and SUNGG galaxies. Along this line, the low F$_{line}$/f$_{FUV}$ ratios of outer-HIIs in conjunction with their relative paucity in comparison with the multitude of GALEX XUV clusters may also be suggestive of a truncated IMF. Detailed calculations of expected F$_{line}$/f$_{FUV}$ and the fraction of \ha-emitting clusters in XUV disks for various star formation histories and stellar IMFs could further support or reject this claim for outer disk star formation.

\section{Properties of the Sample}
\label{properties}

 We show the distribution of ELdot equivalent widths, line fluxes, $R$ Band magnitudes, and r/r$_{25}$ values  in Figures \ref{properties} and \ref{r25area}.  To indicate our source identification, we use a nested shading scheme in Figure \ref{properties}. The clear, outer histogram shows all ELdots in this study, light gray indicates those ELdots we have identified as background galaxies and outer-HIIs via follow-up data (GALEX imaging and optical spectroscopy), and black, the innermost histogram, specifically shows only outer-HIIs. Thus, the visible clear area corresponds to ELdots with no follow-up data, and the visible light gray area indicates background galaxies. 
 \begin{figure}[h!]
\begin{center}
\vspace{-0.01in}
\includegraphics[height=2.4in]{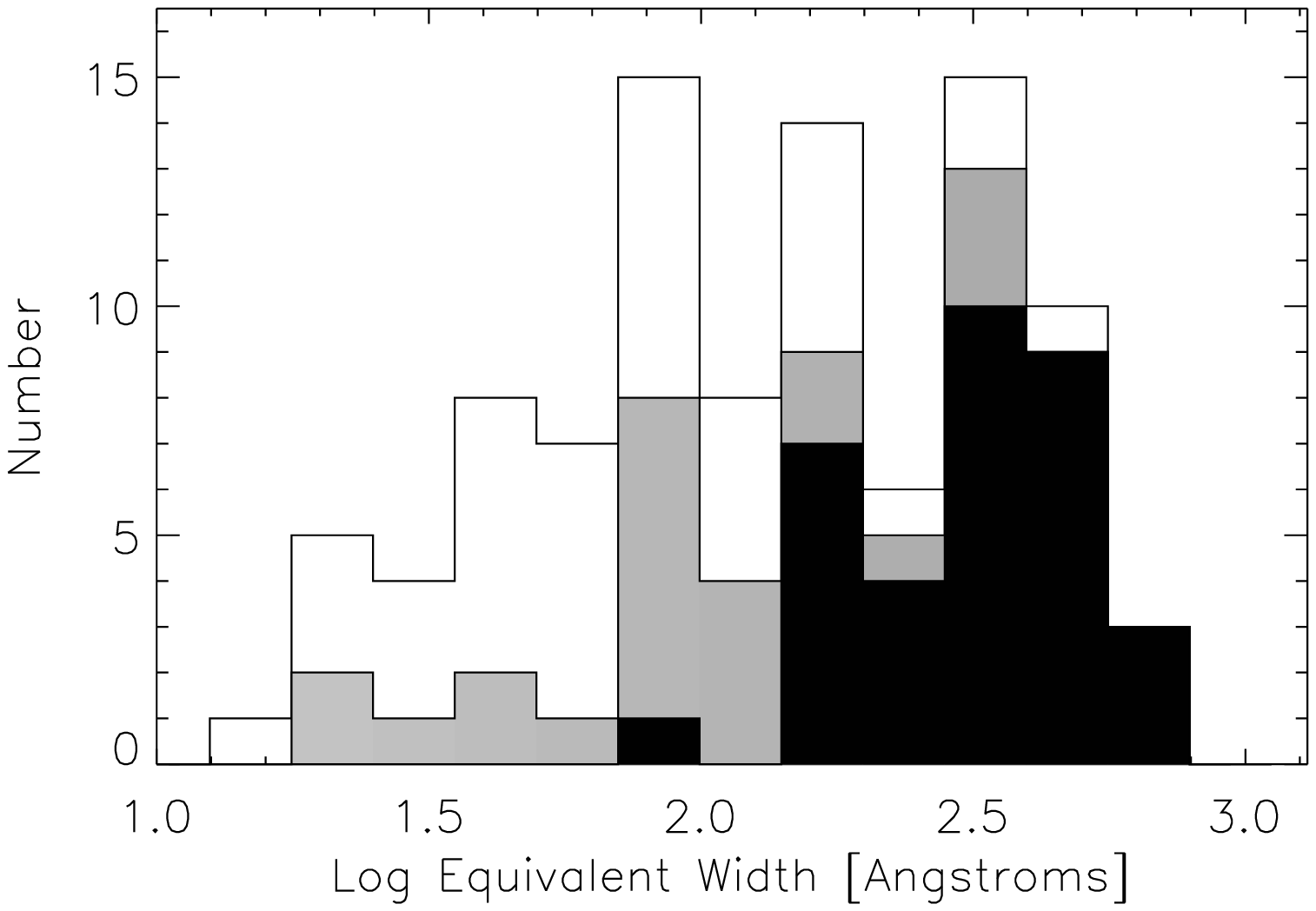}
\includegraphics[height=2.4in]{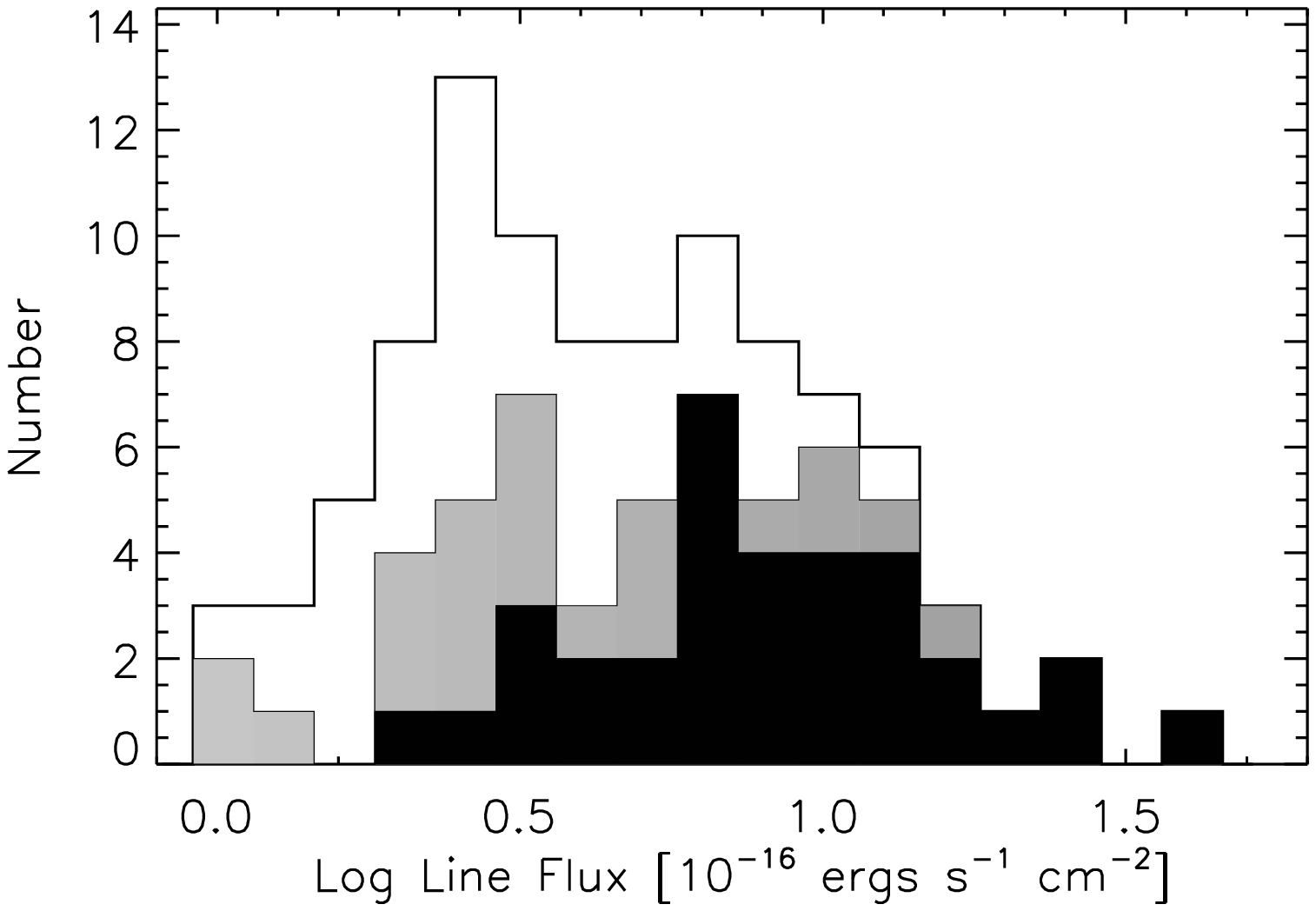}
\includegraphics[height=2.4in]{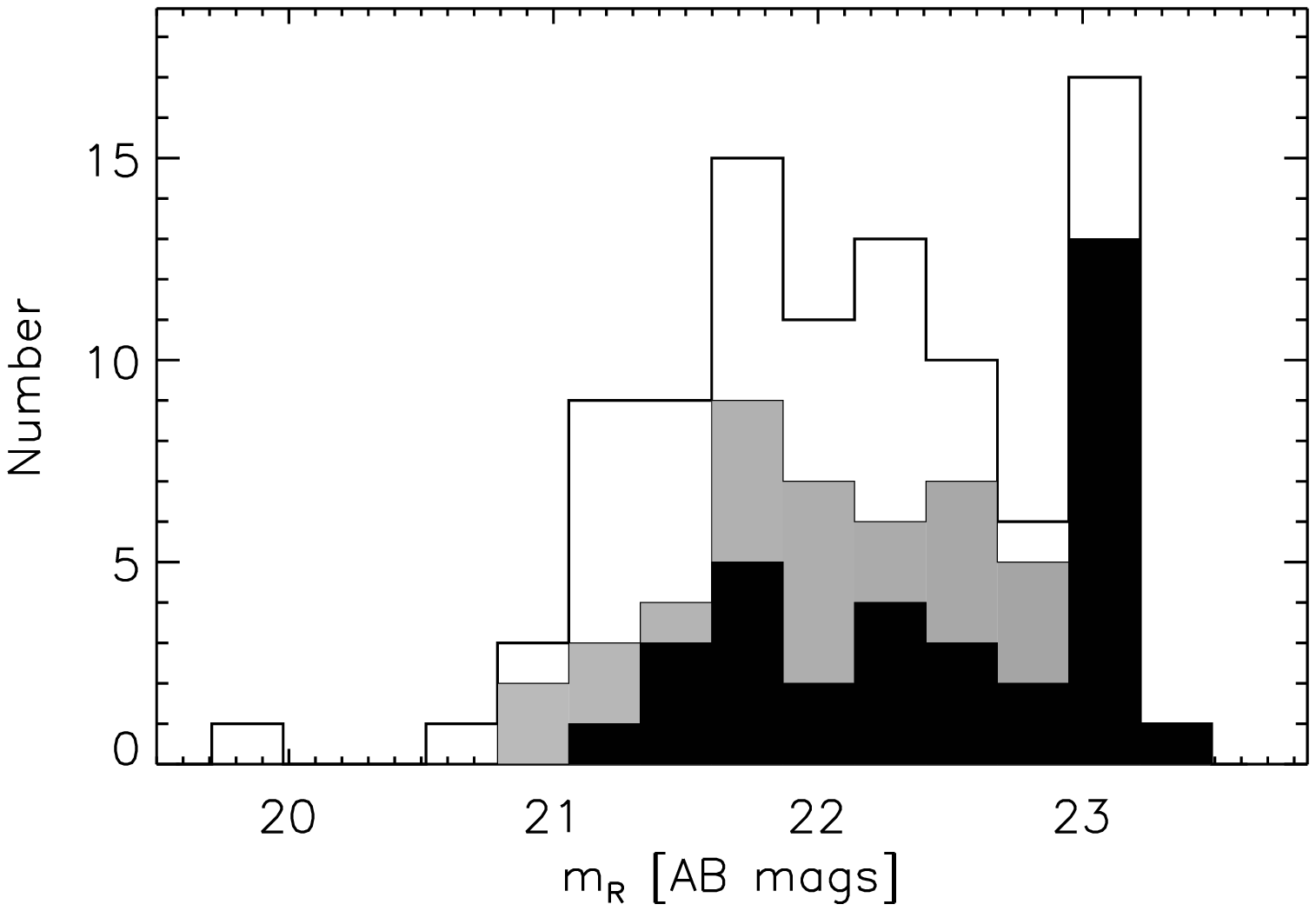}
\end{center}
\caption{\label{properties} Top: Histogram showing the distribution of equivalent widths for the line detected in the SINGG bandpass for all ELdots (clear),  for ELdots with follow-up data (gray; background galaxies plus outer-HIIs) and for outer-HIIs (black). Middle: Histogram showing the distribution of ELdot line fluxes in the SINGG narrow-band filter (\ha~flux, if at the velocity of the target galaxy) with the same shading scheme as above. Bottom: Histogram showing the distribution of SINGG $R$ Band magnitudes for the ELdots, again with the same shading scheme.}
\end{figure}

 To calculate the narrow-band emission's equivalent width in \AA, we divide the flux of the ELdot in the continuum subtracted, net \ha~image by its flux density in the $R$ band. Seen in the top panel of Figure \ref{properties},  the ELdots span a range in EWs from the cut for our sample selection at approximately 20 \AA~to 900 \AA.  Beyond 150 \AA, the majority of ELdots are \ha~emitting outer-HIIs with EW's that extend out to a maximum value of 900 \AA.  The median values of the EW's are 160 \AA~ for the entire sample of ELdots  and 412 \AA~ for the  outer-HIIs.  A simple two-sided KS-test on the distributions of EWs shows a 5\%~ probability that all outer-HIIs are derived from the overall distribution, and a 59\%~ probability for all  background galaxies and outer-HIIs. Therefore,  outer-HIIs tend to have higher EWs than the entire sample of ELdots and known background galaxies. 
 
The distribution of line fluxes in log space is shown in the middle panel of Figure \ref{properties}.  The majority of the ELdots have line fluxes less than $1 \times 10^{-15}$ ergs s$^{-1}$ cm$^{-2}$.  The median  values are $4.93 \times 10^{-16}$ ergs s$^{-1}$ cm$^{-2}$ (entire sample) and $8.9 \times 10^{-16}$ ergs s$^{-1}$ cm$^{-2}$ (outer-HIIs). The limiting \ha~ luminosity of a point source at the median distance of the SINGG sample is $10^{37}$ ergs s$^{-1}$, or  $\sim1.0 \times 10^{-16}$ ergs s$^{-1}$ cm$^{-2}$, roughly the ionizing output of a single O7V star \citep{martins}.  Outer-HIIs tend to have slightly higher line fluxes than the full sample of ELdots. 
  
The distribution of R band magnitudes shown in the bottom panel of Figure \ref{properties} has a more narrow distribution than the line fluxes based on our strict color cut of NB$-$R = -0.7 and the detection limit of the survey.  The mean R band magnitude is 22.15 for the entire sample and 22.48 for the outer-HIIs.   Outer-HIIs comprise the majority of sources at the faint end of this distribution. A two-sided KS-test on the distributions of $R$ band magnitudes gives a 39\% probability that  background galaxies and outer-HIIs are drawn from the full sample and only a 5\% probability that outer-HIIs are. Both fainter $R$ band magnitudes and higher line fluxes contribute to the tendency of outer-HIIs to have higher EWs than the entire sample of ELdots.

 A full understanding of the locations of the ELdots in the SINGG images requires a calculation of the finder search area as a function of  r/r$_{25}$. The top panel of Figure \ref{r25area} shows the total area in elliptical annuli of the 93 SINGG images at given r/r$_{25}$. We calculated this value by summing the area in each elliptical annulus of a given r/r$_{25}$  for every SINGG image. Most SINGG images do not extend beyond 17 r/r$_{25}$, and the majority of the area in each image lies within 5 r/r$_{25}$. This histogram is helpful for understanding those shown in the two lower panels of Figure \ref{r25area}, the distribution of r/r$_{25}$ values for the ELdots. Our shading scheme for Figure \ref{r25area} differs from the nested shading scheme of Figure \ref{properties}. The middle panel of Figure \ref{r25area} shows the distributions of all SINGG ELdots {\emph{minus}} the outer-HIIs (clear) and all  background galaxies (light gray). These distributions roughly follow that of the search area, indicating that background galaxies (which comprise most of the SINGG ELdot sample) are distributed randomly in the SINGG fields. The majority of ELdots are found within 5 $\times$ r/r$_{25}$, because that is where the majority of the search area lies. On the bottom panel of Figure \ref{properties}, we show the distribution of  r/r$_{25}$ values for the outer-HIIs. This histogram includes the 23 outer-HIIs of J0403-43 all within the first bin, and thus is heavily biased for that system. Nonetheless, outer-HIIs appear more likely to be found closer to a host galaxy than background galaxies. A two-sided KS test indicates a 98\% probability that background galaxies and the full sample of ELdots come from the same distribution, while the probability is 24\% that the outer-HIIs come from the same distribution. 
\begin{figure}[h!]
\epsscale{1.3}
\plotone{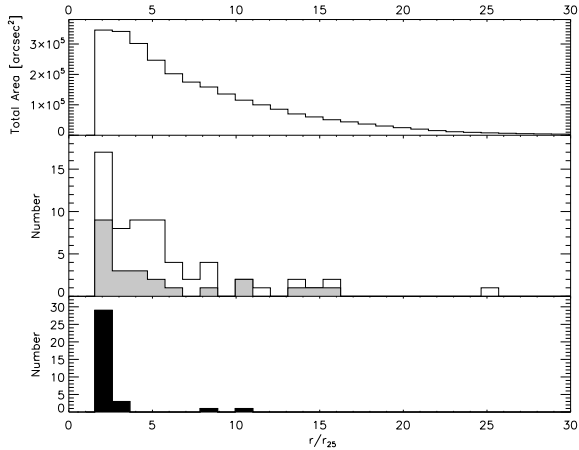}
\figcaption{\label{r25area} Top: The total area in square arcseconds probed in elliptical annuli at given r/r$_{25}$ for all 93 SR1 fields. Middle: the distribution of  r/r$_{25}$ values for all SINGG ELdots minus  outer-HIIs (clear) and background galaxy ELdots (gray). Bottom: the distribution of r/r$_{25}$ values for SINGG outer-HIIs. }
\end{figure}

\section{Discussion of Outer-HIIs}
\label{disc}

The outlying HII regions discussed here are not of a uniform nature, nor are they found in a set of uniform systems.  Some are OB associations, isolated in the far outskirts of galaxies, others are part of extended UV disks, and others may be dwarf satellite companions to host SINGG galaxies. Furthermore, outer-HIIs are found in systems that span a large range of  HI mass probed by SINGG (M06), from Log M$_{HI}$ = 8.25 (J0019-22) to 10.31 (J0209-10a). What links all of the outer-HIIs together is that they represent new stars forming in the absence of a significant, if any, underlying stellar population. As we have discussed, part of the variety in the outer-HIIs arises because of the range of distances probed by the SINGG survey, from 4 Mpc to 70 Mpc. Thus, compact objects roughly the angular size of the SINGG  PSF of 1.6$\arcsec$ will span a range of physical sizes between 30 and 500 parsecs, as discussed in Section \ref{selectioneffects}. Those physical sizes correspond to compact star clusters in the most nearby cases, to small dwarf galaxies in the most distant systems. Conversions of \ha~fluxes to \ha~luminosities use distances listed in table 6 of M06, except for NGC 1533 (J0409-56) which has a distance of 19.4 Mpc determined by \cite{deGraaf07}. For conversions to an equivalent number of O9V stars, we use Q$_{o}$ = 7.33 $\times$ 10$^{11}$ L$_{H\alpha}$ s$^{-1}$ \citep{osterbrock} and the ionizing photon output of a single O9V star to be 10$^{47.90}$ s$^{-1}$ \citep{martins}. The seven SINGG systems containing outer-HIIs are presented below in order of SINGG target galaxy distance: 
 
 \begin{enumerate} 
\item{ {\bf{J0019-22:}} The two outer-HIIs are young star clusters or OB associations located in a Type 1 XUV disk around the host dwarf galaxy, MCG 04-02-003. A distance of 10 Mpc gives maximum physical diameters of the HII regions of 70 pc, roughly in line with those of the resolved stellar clusters associated with the outer-HIIs outside of NGC 1533 and similar in size to Galactic OB associations \citep{werk08}. The environment of these outer-HIIs however, is distinct from that of NGC 1533 in that they seem to fit in a faint, organized disk. The outer-HIIs lie at projected galactocentric distances of roughly 5 kpc, on the edges of the XUV disk.The \ha~ luminosities of the two outer-HIIs are similar, roughly 10$^{36.6}$ ergs s$^{-1}$, the equivalent of about 4 O9V stars. J0019-22 is a solitary galaxy, though it is slightly peculiar in that it possesses a star-forming compact core and an optically very faint but FUV-bright XUV disk. We note that this galaxy is the only solitary SINGG galaxy that contains  outer-HIIs based on our morphological classification system.}

\item{{\bf{J0403-43:}} The numerous outer-HIIs associated with the NGC 1512/10 pair are organized in outer spiral features that align quite well with the XUV GALEX complexes. A distance of 10.4 Mpc \citep{gildepaz07} gives a maximum physical diameter for these outer-HIIs of 75 pc, again, in line with typical Galactic OB associations. Projected galactocentric distances of the outer-HIIs range from 15 - 20 kpc, while the FUV-bright spiral arms extend beyond the SINGG image, to greater than 30 kpc in projected distance from the center of NGC 1512.  The \ha~luminosities span a rather broad range, between 10$^{36.9}$ and 10$^{37.9}$ ergs s$^{-1}$, representing anywhere between 5 and 50 O9V stars. The NGC 1512/10 pair is a nearby interacting galaxy pair with an Type 1 XUV disk (see Thilker et al. 2007). }

\item{{\bf{J0355-42:}} Here, we see a single outer-HII that is part of the extended, young star formation triggered by a violent galaxy collision that formed the galaxy NGC 1487 \citep{mengel08}. The H$\alpha$ emission is coming from the tip of a much larger FUV-bright stellar complex at the outskirts of NGC 1487, roughly 13 kpc in projection from the system's center. In the GALEX image shown in Figure \ref{ngc1487}, we can see that this complex may be part of a faint arm-like structure extending from the central, optically bright part of NGC 1487. There are two additional  high-EW point sources along this faint structure (seen in the SINGG H$\alpha$ image, also with FUV counterparts), though they lie within 2 $\times$ r/r$_{25}$. The physical size ($\sim80$ pc at D = 11.2 Mpc) and H$\alpha$ luminosity (10$^{36.7}$ ergs s$^{-1}$) of the outer-HII are again consistent with that of a typical OB association. }

\item{ {\bf{J0409-56:}} The 4 outer-HIIs outside NGC 1533 lie in a ring of  HI gas and are powered by young, low-mass OB associations revealed by $\it{HST}$ HRC images presented in \cite{werk08}.  \ha~luminosities between 10$^{37.4}$ and 10$^{37.7}$ ergs s$^{-1}$ correspond to emission from 20 -40 O9V stars. The GALEX image of the system (Figure \ref{ngc1533galex}) shows that these OB associations are part of recent extended star formation as traced by its Type 1 XUV emission, represented by a number of FUV-bright stellar complexes in the southeast part of the HI ring. The overdensity of FUV sources here fits well with the picture presented by \cite{werk08} that included a number of blue field objects that did not show any \ha~ emission. The HI ring is connected to a companion galaxy  in the upper-right of Figure \ref{ngc1533galex} (see \cite{emmaconf} for more details), and we therefore classify NGC 1533 as interacting. }

\item{{\bf{J2202-20:}} The outer-HII near the edge-on spiral galaxy NGC 7184 is an \ha-emitting knot in an extension of the galaxy's optical disk (see Figure \ref{j2202-20_eldots}). The galaxy's distance of 37 Mpc gives the outer-HII a maximum physical FWHM size of 270 pc, comparable to the star-forming complexes present in the XUV disks of the GALEX sample \citep{Thilker07}.  This outer-HII lies on the northeast tip the rotating HI-disk of NGC 7184.  Its \ha~luminosity,  10$^{37.6}$ ergs s$^{-1}$, is equivalent to that from 35 O9V stars. While NGC~7184 is the dominant galaxy in the J2202$-$20 field, there is a dwarf galaxy to the south east which is also a line emitter as identified by M06. It is not clear whether these two galaxies are interacting, and so we have designated this system as one with a companion. }

\item{{\bf{J0317-22:}} The outer-HII may be a dwarf galaxy near the spiral ESO481-G017 given its large velocity offset from the main galaxy and their distance of 52 Mpc, which places the maximum size of the outer-HII at nearly 400 pc. The projected distance from the galaxy center to the outer-HII is 43 kpc.  The \ha~ luminosity is  considerably higher than more nearby outer-HIIs (10$^{38.1}$ ergs s$^{-1}$) and is equivalent to 125 O9V stars. High-resolution HI observations of the system arepresented by \cite{nitza09}, and show ESO481-G017 to have several companions. }

\item{ {\bf{J0209-10a:}} The two outer-HIIs lie in the intracluster medium of Hickson Compact Group (HCG) 16. Because this system lies at a distance of 54 Mpc, these two outer-HIIs are probably not simply single isolated star clusters, but could be complexes of  young star clusters or emerging dwarf galaxies. As with J0317-22:E1, the \ha~luminosities of the two outer-HIIs range from 10$^{38.2}$ and 10$^{38.5}$ ergs s$^{-1}$, which translates to hundreds of O9V stars. The outer-HIIs lie in a large extension of HI gas stripped from the galaxies in the group \citep{verdesmontenegro01} due to the ongoing interaction in HCG 16. }

\end{enumerate}

Examples in the existing scientific literature of objects similar to the outer-HIIs presented here tend to showcase interacting systems and galaxy groups as the locations for outer-galaxy or intracluster star formation \citep{boquien07, walter, oosterloo, cortese, sakai, gerhard, arnaboldi}. Furthermore, the incidence of Type-1 XUV disks in GALEX is shown to correlate with galaxy interactions \citep{Thilker07}. We can confirm that outer-HIIs tend to be found in interacting systems or systems with companions. Six of the seven systems presented here exhibit such a nature. In contrast, the majority of the SR1 systems (and systems containing background galaxy ELdots) are solitary. Additionally, the SINGG fields containing outer-HIIs that have available long-exposure GALEX data (J0409-56, J0019-22, J0403-43, and J0355-42) show that outer-HIIs are strongly related to extended UV emission, each associated with Type 1 XUV morphologies.  In these cases, the \ha~emission traces only the youngest, most massive clusters among a more extensive population of star clusters in the low density outskirts of galaxies. We will examine the connection between outer-HIIs and XUV emission in a future paper. Of particular interest is a calculation of the number of expected \ha-emitting massive O stars at large galactocentric radii given the luminosity of the XUV disk and assuming a Salpeter stellar IMF.  

 We can estimate the frequency of outer-HIIs in gas-rich galaxies with our results. Of the 33 systems for which we have follow-up data (optical spectra: 26 systems; GALEX deep images: 7 additional systems),  seven contain outer-HIIs, or $\sim20$\%. There are 17 distinct systems with ELdots for which we have no spectroscopic or GALEX follow-up data. On the whole, the ELdots in these systems have lower equivalent widths and brighter R-band magnitudes than the  outer-HIIs (see Figure \ref{properties}).  If, at most, $\sim$20\% of these 17 systems lacking follow-up contain outer-HIIs, then we would expect, at most,  3 more systems with outer-HIIs. For the overall frequency of outer-HIIs in the entire SR1 sample, we include the additional 39 systems in which there are no finder-detected ELdots. The frequency of star formation beyond 2 $\times$ r$_{25}$ in an unbiased sample of 89 gas-rich systems spanning a wide range of morphologies and masses is therefore between 8 and 11\% (7 - 10 systems).

As noted in Section 3.3, outer-HIIs with distances beyond 30 Mpc likely represent emission from multiple HII regions, and perhaps potential dwarf galaxies. Comparatively, the outer-HIIs with distances less than 30 Mpc  are more similar in size and luminosity to HII regions powered by single OB associations, and are seen to be part of extended UV emission associated with the host galaxy. As there is much recent interest in the latter type of outer-HII, we additionally calculate the frequency of outer-HIIs in SINGG galaxies with D $<$ 30 Mpc (62 of 89 SR1 systems).  A total of 32 of these systems contain ELdots, though only 22 have follow up data (optical spectra: 16 systems; GALEX deep images: 6 additional systems). 4 of these 22 systems contain outer-HIIs, or $\sim20$\%. Similar to the calculation performed above, we then expect, at most, 2 of the additional 10 systems with ELdots but without follow-up data to contain outer-HIIs. Thus, the frequency of outer-HIIs in our distance-limited sample is between 6 and 10\% (4 - 6 systems).

\section{Summary and Conclusions}
\label{conc}

	ELdots are emission-line point sources (``dots") well outside the main optical R-band emission of a galaxy, defined as 2 $\times$ r$_{25}$ in this work. Using an automated finder, we catalogue a total of 96 ELdots in 50 of 89 SINGG systems.   We classify ELdots as either outlying HII regions (outer-HIIs) or background galaxies by spectroscopy and GALEX morphology and colors. This study highlights four key results: 
	\begin{itemize}
	
	\item{Follow-up GALEX data, combined with SINGG photometric properties, can help distinguish between outer-HIIs and background galaxy ELdots, nearly eliminating the need for spectroscopic follow-up. The distribution of ELdot FUV$-$R colors is bimodal, revealing that outer-HIIs are considerably bluer than higher-redshift emission-line background galaxies. The very blue outer-HII colors are due primarily to the extremely low R-band emission from these objects, indicating new stars forming where there is little or no underlying older stellar population. Outer-HIIs also tend to have higher emission-line EWs than background galaxy ELdots and are likely to be found closer to a potential host galaxy than the background galaxies, which are uniformly distributed across the SINGG search area.}
	
	 \item{Through our systematic search of a sample of gas-rich galaxies, we have confirmed that massive star formation in the far outskirts of galaxies tends to be associated with galaxy interactions and nearby companions. This result agrees with other recent findings of star formation outside the main optical bodies of galaxies. }

\item{ In the cases where long-exposure GALEX data is available, we find that outer-HIIs appear to be associated with XUV emission, specifically of Type-1 morphology.  }

	\item{ We find that the frequency of massive star formation in the far outskirts of galaxies, as traced by \ha~emission in the full SR1 sample, is between 8 and 11\%. Outer-HII regions in this full sample span a wide range of H$\alpha$ luminosities, from 10$^{36.6}$ ergs s$^{-1}$ to 10$^{38.5}$ ergs s$^{-1}$, and have physical sizes from 70 pc to 500 pc, representing OB associations ionized by a few O stars in the most nearby systems, and dwarf galaxies with hundreds of O stars in the most distant systems. When we isolate our sample of outer-HIIs to a nearby sample with D $<$ 30 Mpc, we can compare our frequency statistics more directly with those of XUV galaxies. These outer-HIIs are more homogeneous in nature, similar in size and luminosity to Galactic OB associations. Where \cite{Thilker07} find XUV emission to be present in $\sim30$\% of spiral galaxies in the local universe, we find the frequency of outlying HII regions in gas-rich galaxies more nearby than 30 Mpc to be between 6 and 10\%. }

\end{itemize}

\section{Acknowledgements}

JKW acknowledges support from an NSF Graduate Research Fellowship, and very useful conversations with David Schiminovich, Nitza Santiago, Jennifer Donovan, Jacqueline van Gorkom, and Joo Heon Yoon. JKW also enthusiastically thanks the Columbia Astronomy and Astrophysics Department for welcoming her as a full-time visiting student beginning in the Fall of 2008. Follow-up spectroscopic observations were supported in part by Spitzer Program ID 20695. {\emph{Spitzer Space Telescope}} is operated by the Jet Propulsion Laboratory, California Institute of Technology, under a contract with NASA.  MEP acknowledges additional support from NSF AST-0904059. This research has made use of the NASA/IPAC Extragalactic Database (NED) which is operated by the Jet Propulsion Laboratory, California Institute of Technology, under contract with the National Aeronautics and Space Administration.

\bibliography{references}
\bibliographystyle{apj}

%%%%%%%%%%%%%%%%%%%%%%%%%%%%%%%%%%%%%%%%%%%%%%%%%%%%%%%%%%%%%%%%%%%%

%\theendnotes
%%%%%%%%%%%%%%%%%%%%%%%%%%%%%%%%%%%%%%%%%%%%%%%%%%%%%%%%%%%%%%%%%%%%

\clearpage

%%%%%%%%%%%%%%%%%%%%%%%%%%%%%%%%%%%%%%%%%%%%%%%%%%%%%%%%%%%%%%%%%%%%%
\clearpage

%\begin{landscape}
\begin{deluxetable}{l c c c c c c c c c c}
  \tablecaption{ELdots Found in SINGG Images \label{t:sr1prop}}
  \tabletypesize{\scriptsize}
%  \rotate
  \tablewidth{0pt}
  \tablehead{\colhead{Designation} &
             \colhead{RA} &
             \colhead{dec} &
             \colhead{EW} &
             \colhead{r/r$_{25}$} &
             \colhead{F$_{Line}$} &
             \colhead{m$_{R}$} &
             \colhead{Run ID} &
             \colhead{Line Obs} &
             \colhead{Morphology}\\
             \colhead{(1)} &
             \colhead{(2)} &
             \colhead{(3)} &
             \colhead{(4)} &
             \colhead{(5)} &
             \colhead{(6)} &
             \colhead{(7)} &
             \colhead{(8)} &
             \colhead{(9)} &
             \colhead{(10)}}
            
\startdata
J0005-28:E1 &   00~05~19.0 &    -28~10~24 &   103 & 15.9  &  2.74$\pm$0.23 & 22.44$\pm$0.168 &  Nov05 &    [OII] &      int &   \\
J0019-22:E1 &   00~19~10.5 &    -22~41~25 &   263 &  2.5  &  3.97$\pm$0.37 & 23.06$\pm$0.265 &  ..... &     .... &     solo &   \\
J0019-22:E2 &   00~19~11.1 &    -22~38~35 &   299 &  2.7  &  3.90$\pm$0.37 & $>$23.21 &  ..... &     .... &       "" &   \\
J0031-22:E1 &   00~31~12.4 &    -22~53~20 &    30 & 16.6  &  2.64$\pm$0.39 & 21.17$\pm$0.058 &  ..... &     .... &     solo &   \\
J0031-22:E2 &   00~31~34.2 &    -22~49~13 &    27 & 14.2  &  2.12$\pm$0.39 & 21.27$\pm$0.063 &  ..... &     .... &       "" &   \\
J0039-14:E1 &   00~39~00.2 &    -14~13~09 &    58 &  3.9  &  3.16$\pm$0.25 & 21.67$\pm$0.088 &  ..... &     .... &      int &   \\
J0039-14:E2 &   00~39~19.1 &    -14~08~04 &    53 &  4.9  &  4.21$\pm$0.27 & 21.26$\pm$0.063 &  ..... &     .... &       "" &   \\
J0156-68:E1 &   01~55~36.8 &    -69~03~08 &    42 & 25.9  &  2.85$\pm$0.33 & 21.43$\pm$0.071 &  ..... &     .... &     solo &   \\
J0209-10a:E1 &   02~09~16.2 &    -10~13~10 &   148 &  6.6  &  9.34$\pm$1.16 & 21.48$\pm$0.095 &  ..... &     .... &      int &   \\
J0209-10a:E2 &   02~09~27.6 &    -10~07~16 &   456 &  8.4  &  8.14$\pm$1.15 & $>$22.85 &  Sep02 &      \ha &       "" &   \\
J0209-10a:E3 &   02~09~37.6 &    -10~05~35 &   191 & 10.9  &  5.00$\pm$1.12 & 22.43$\pm$0.212 &  Nov05 &      \ha &       "" &   \\
J0221-05:E1 &   02~21~49.5 &    -05~27~41 &   241 &  3.5  &  4.47$\pm$0.88 & 22.82$\pm$0.221 &  ..... &     .... &      int &   \\
J0221-05:E2 &   02~21~41.2 &    -05~26~29 &    90 &  3.7  &  3.60$\pm$0.87 & 21.99$\pm$0.110 &  Nov05 &   [OIII] &       "" &   \\
J0221-05:E3 &   02~21~29.7 &    -05~23~50 &    80 &  5.3  &  4.93$\pm$0.88 & 21.52$\pm$0.074 &  ..... &     .... &       "" &   \\
J0223-04:E1 &   02~23~50.7 &    -04~37~55 &   630 &  2.4  &  9.62$\pm$0.68 & 23.03$\pm$0.247 &  ..... &     .... &     solo &   \\
J0223-04:E2 &   02~23~53.0 &    -04~37~28 &   213 &  2.2  &  6.30$\pm$0.62 & 22.31$\pm$0.136 &  ..... &     .... &       "" &   \\
J0223-04:E3 &   02~23~54.3 &    -04~37~05 &   387 &  2.3  &  9.15$\pm$0.67 & 22.55$\pm$0.166 &  ..... &     .... &       "" &   \\
J0224-24:E1 &   02~25~21.8 &    -24~51~59 &    84 &  6.2  &  3.69$\pm$0.63 & 21.87$\pm$0.087 &  ..... &     .... &      int &   \\
J0256-54:E1 &   02~56~38.6 &    -54~40~35 &    59 &  3.8  &  2.71$\pm$0.91 & 21.86$\pm$0.112 &  ..... &     .... &     solo &   \\
J0256-54:E2 &   02~57~17.3 &    -54~39~16 &    27 &  8.1  &  1.80$\pm$0.91 & 21.47$\pm$0.081 &  ..... &     .... &       "" &   \\
J0256-54:E3 &   02~56~31.3 &    -54~31~35 &    54 &  5.2  &  2.94$\pm$0.92 & 21.67$\pm$0.096 &  ..... &     .... &       "" &   \\
J0317-22:E1 &   03~17~10.3 &    -22~54~18 &   331 &  2.2  &  3.70$\pm$0.64 & $>$23.35 &  Nov05 &      \ha &     comp &   \\
J0317-22:E2 &   03~16~59.3 &    -22~49~39 &   132 &  2.3  &  2.44$\pm$0.64 & 22.81$\pm$0.194 &  Nov05 &   [OIII] &       "" &   \\
J0320-52:E1 &   03~19~49.2 &    -52~05~24 &   206 &  8.7  &  7.22$\pm$0.40 & 22.14$\pm$0.124 &  ..... &     .... &     solo &   \\
J0322-04:E1 &   03~23~12.0 &    -04~15~16 &   332 &  8.5  &  12.0$\pm$0.58 & 22.09$\pm$0.110 &  Oct03 &   [OIII] &     solo &   \\
J0322-04:E2 &   03~22~54.3 &    -04~12~58 &    91 &  3.6  &  6.06$\pm$0.44 & 21.43$\pm$0.064 &  ..... &     .... &       "" &   \\
J0322-04:E3 &   03~22~50.2 &    -04~10~02 &    74 &  2.7  &  5.40$\pm$0.43 & 21.32$\pm$0.058 &  ..... &     .... &       "" &   \\
J0322-04:E4 &   03~22~53.1 &    -04~09~26 &    64 &  3.7  &  2.01$\pm$0.39 & 22.24$\pm$0.126 &  ..... &     .... &       "" &   \\
J0341-01:E1 &   03~41~30.5 &    -01~58~09 &    54 &  8.9  &  2.31$\pm$0.25 & 21.91$\pm$0.084 &  ..... &     .... &     solo &   \\
J0348-39:E1 &   03~48~22.3 &    -39~27~03 &   121 &  4.3  &  3.57$\pm$1.20 & 22.32$\pm$0.205 &  ..... &     .... &     solo &   \\
J0351-38:E1 &   03~51~55.1 &    -38~29~22 &    70 &  6.2  &  2.97$\pm$0.36 & 21.94$\pm$0.105 &  ..... &     .... &     solo &   \\
J0355-42:E1 &   03~55~59.1 &    -42~23~03 &   206 &  2.2  &  2.89$\pm$0.45 & $>$23.14 &  ..... &     .... &      int &   \\
J0359-45:E1 &   03~58~47.9 &    -45~47~08 &    43 &  8.0  &  1.84$\pm$0.73 & 21.95$\pm$0.118 &  ..... &     .... &     comp &   \\
J0409-56:E1 &   04~10~13.6 &    -56~11~37 &   677 &  3.7  &  1.05$\pm$0.63 & $>$23.02 &  Sep02 &      \ha &      int &   \\
J0409-56:E2 &   04~10~14.4 &    -56~11~35 &   456 &  3.7  &  7.07$\pm$0.57 & $>$23.02 &  Sep02 &      \ha &       "" &   \\
J0409-56:E3 &   04~10~16.8 &    -56~10~46 &   412 &  3.4  &  6.39$\pm$0.57 & $>$23.02 &  IMACS &      \ha &       "" &   \\
J0409-56:E4 &   04~10~15.6 &    -56~06~14 &   318 &  2.1  &  4.93$\pm$0.56 & $>$23.02 &  Sep02 &      \ha &       "" &   \\
J0441-02:E1 &   04~41~12.9 &    -02~53~46 &   176 &  2.4  &  3.60$\pm$0.54 & 22.73$\pm$0.265 &  Dec06 &   [OIII] &     solo &   \\
J0457-42:E1 &   04~56~38.9 &    -42~43~26 &    16 & 12.5  &  1.25$\pm$0.33 & 21.30$\pm$0.064 &  ..... &     .... &     solo &   \\
J0459-26:E1 &   05~00~16.4 &    -25~59~12 &   374 &  2.6  &  9.16$\pm$1.23 & 22.53$\pm$0.250 &  Nov05 &   [OIII] &     solo &   \\
J0503-63:E1 &   05~03~24.9 &    -63~42~59 &    87 &  2.7  &  7.74$\pm$0.81 & 21.11$\pm$0.072 &  ..... &     .... &     comp &   \\
J0504-16:E1 &   05~04~28.0 &    -16~35~55 &   430 &  2.6  &  12.0$\pm$1.08 & 22.36$\pm$0.162 &  ..... &     .... &      int &   \\
J0506-31:E1 &   05~05~57.7 &    -31~55~05 &    28 &  5.0  &  2.28$\pm$0.48 & 21.23$\pm$0.061 &  Dec06 &   [OIII] &     solo &   \\
J0507-16:E1 &   05~07~42.5 &    -16~21~43 &   102 &  2.4  &  2.47$\pm$0.84 & 22.53$\pm$0.169 &  Nov05 &   [OIII] &     solo &   \\
J0507-37:E1 &   05~07~29.8 &    -37~23~24 &   182 &  3.2  &  18.2$\pm$0.90 & 21.00$\pm$0.055 &  Oct03 &   [OIII] &     solo &   \\
J0510-31:E1 &   05~11~08.7 &    -31~42~51 &   108 &  6.3  &  1.37$\pm$0.34 & $>$23.25 &  ..... &     .... &     solo &   \\
J0510-31:E2 &   05~11~15.6 &    -31~41~17 &   187 &  5.8  &  2.67$\pm$0.35 & 23.11$\pm$0.272 &  ..... &     .... &       "" &   \\
J0512-32:E1 &   05~11~28.1 &    -32~55~41 &   133 &  5.0  &  10.3$\pm$1.13 & 21.28$\pm$0.075 &  Oct03 &   [OIII] &     solo &   \\
J0943-09:E1 &   09~43~42.5 &    -10~00~55 &   342 &  7.1  &  5.79$\pm$0.61 & 22.91$\pm$0.245 &  May06 &   [OIII] &     solo &   \\
J1002-05:E1 &   10~02~46.9 &    -05~55~01 &    28 &  5.9  &  7.15$\pm$0.47 & 19.99$\pm$0.028 &  May06 &     star &     solo &   \\
J1018-17:E1 &   10~18~06.9 &    -18~00~04 &   123 &  9.3  &  3.94$\pm$0.79 & 22.22$\pm$0.142 &  ..... &     .... &     solo &   \\
J1051-19:E1 &   10~25~17.3 &    -20~03~28 &   138 & 11.0  &  6.23$\pm$0.56 & 21.85$\pm$0.094 &  May06 &   [OIII] &     solo &   \\
J1051-19:E2 &   10~25~14.7 &    -20~03~09 &    49 & 11.1  &  1.23$\pm$0.51 & 22.49$\pm$0.162 &  Dec06 &   H$\beta$ &       "" &   \\
J1054-18:E1 &   10~54~25.4 &    -18~09~17 &   161 &  6.5  &  12.7$\pm$0.59 & 21.24$\pm$0.046 &  ..... &     .... &     comp &   \\
J1109-23:E1 &   11~10~06.1 &    -23~42~13 &    22 &  2.2  &  1.07$\pm$0.90 & 21.81$\pm$0.090 &  May06 &   [OIII] &     solo &   \\
J1131-02:E1 &   11~31~23.1 &    -02~23~15 &    80 &  5.3  &  2.95$\pm$0.66 & 22.07$\pm$0.133 &  May06 &   [OIII] &     comp &   \\
J1131-02:E2 &   11~31~40.4 &    -02~16~45 &    79 &  2.8  &  2.86$\pm$0.66 & 22.09$\pm$0.136 &  May06 &   [OIII] &       "" &   \\
J1341-29:E1 &   13~41~53.0 &    -29~58~25 &   105 &  2.8  &  4.77$\pm$1.18 & 21.86$\pm$0.159 &  May06 &   [OIII] &      int &   \\
J2009-61:E1 &   20~08~43.1 &    -61~57~26 &   112 & 15.2  &  3.19$\pm$0.57 & 22.37$\pm$0.168 &  May06 &   [OIII] &     solo &   \\
J2052-69:E1 &   20~52~59.3 &    -69~12~26 &   149 &  4.5  &  2.96$\pm$0.50 & $>$22.76&  May06 &   [OIII] &     solo &   \\
J2102-16:E1 &   21~02~19.4 &    -16~50~11 &   103 &  2.1  &  3.61$\pm$0.36 & 22.14$\pm$0.150 &  Oct03 &   H$\beta$ &     solo &   \\
J2149-60:E1 &   21~48~55.7 &    -60~39~52 &    29 &  2.7  &  1.39$\pm$0.47 & 21.80$\pm$0.095 &  May06 &   [OIII] &      int &   \\
J2202-20:E1 &   22~03~04.1 &    -20~45~48 &    82 &  2.4  &  2.33$\pm$0.32 & 22.36$\pm$0.144 &  May06 &      \ha &     comp &   \\
J2202-20:E2 &   22~02~44.3 &    -20~43~09 &    88 &  4.7  &  1.79$\pm$0.31 & 22.72$\pm$0.196 &  ..... &     .... &       "" &   \\
J2214-66:E1 &   22~15~49.0 &    -66~44~30 &    71 &  4.3  &  2.23$\pm$0.47 & $>$22.26 &  ..... &     .... &     solo &   \\
J2217-42:E1 &   22~16~59.0 &    -42~45~59 &    81 &  5.7  &  2.80$\pm$0.99 & 22.15$\pm$0.119 &  ..... &     .... &     solo &   \\
J2220-46:E1 &   22~21~18.6 &    -46~09~31 &    56 &  4.0  &  5.24$\pm$0.69 & 21.09$\pm$0.060 &  Nov05 &   [OIII] &     solo &   \\
J2234-04:E1 &   22~34~39.6 &    -04~35~53 &   253 & 13.9  &  15.1$\pm$0.65 & 21.57$\pm$0.095 &  Nov05 &   [OIII] &     solo &   \\
J2257-41:E1 &   22~57~51.2 &    -42~50~57 &   114 &  6.2  &  1.77$\pm$0.65 & 23.03$\pm$0.259 &  ..... &     .... &     solo &   \\
J2336-37:E1 &   23~35~50.4 &    -37~59~17 &    39 &  4.6  &  1.72$\pm$0.64 & 21.91$\pm$0.111 &  ..... &     .... &     comp &   \\
J2336-37:E2 &   23~36~29.4 &    -37~58~54 &    46 &  2.3  &  1.33$\pm$0.64 & 22.35$\pm$0.162 &  ..... &     .... &       "" &   \\
J2352-52:E1 &   23~51~51.3 &    -52~34~33 &    33 &  5.4  &  4.44$\pm$0.38 & 20.69$\pm$0.042 &  HST04 &  bkgdgal &      int &   \\
 \enddata
\tablecomments{Column descriptions [units]: (1) Source name.  (2) Source Right Ascension [J2000 hms] (3) Source declination [J2000 dms]. (4) Equivalent Width [$\AA$] (5) Number of times beyond the 25$^{th}$ magnitude elliptical isophote the source lies (6) Emission-line flux of the source measured from SINGG image [ 10$^{-16}$~ergs s$^{-1}$ cm$^{-2}$] (7) R-band continuum magnitude measured from SINGG image [AB mags]  (8) Observing run identification for spectroscopic follow-up, if applicable. HST04 is imaging with the high resolution channel, and in the case of J2352-52, revealed an obvious background galaxy. Exposure times and HST run information are given in \cite{werk08}. (9) The emission-line spectroscopically confirmed to fall within the SINGG narrow-band. \ha~indicates star-formation associated with the SINGG galaxy (outer-HII); [OIII] falling within the \ha~bandpass indicates a source redshift of approximately 0.3; [OII] indicates a redshift of 0.76; and H$\beta$ indicates a redshift of $\sim0.4$; (10) morphological classification described in Section 3.  }
%\tablecomments{{\it Sample portion of table.}}
\end{deluxetable}
%\end{landscape}

%%%%%%%%%%%%%%%%%%%%%%%%%%%%%%%%%%%%%%%%%%%%%%%%%%%%%%%%%%%%%%%%%%%%%
\clearpage

%\begin{landscape}
\begin{deluxetable}{l c c c c c c c c c}
  \tablecaption{Properties of the ELdots in J0403-43 \label{ngc1512tab}}
  \tabletypesize{\scriptsize}
%  \rotate
  \tablewidth{0pt}
  \tablehead{\colhead{Designation} &
             \colhead{RA} &
             \colhead{dec} &
             \colhead{EW} &
             \colhead{r/r$_{25}$} &
             \colhead{F$_{Line}$} &
             \colhead{m$_{R}$} &
             \colhead{FUV} &
             \colhead{FUV - NUV} \\
             \colhead{(1)} &
             \colhead{(2)} &
             \colhead{(3)} &
             \colhead{(4)} &
             \colhead{(5)} &
             \colhead{(6)} &
             \colhead{(7)} &
             \colhead{(8)} &
             \colhead{(9)} }
            
\startdata

J0403-43:E1 &        04~04~20.9  & -43~27~36.4 &   349 &  2.8  &  16.7$\pm$0.86 & 21.80$\pm$0.108  &20.90$\pm$0.07 & -0.06$\pm$0.14\\
J0403-43:E2 &        04~04~21.1  & -43~27~36.6 &   389 &  2.8  &  14.5$\pm$0.80 & 22.08$\pm$0.137 &.....&  .... \\
\hline
J0403-43:E3 &        04~04~32.8  & -43~25~21.1 &   507 &  2.8  &  29.3$\pm$1.24 & 21.60$\pm$0.091 &19.24$\pm$0.06&  -0.22$\pm$0.10 \\
J0403-43:E4 &        04~04~32.9  & -43~25~25.0 &   175 &  2.8  &  8.06$\pm$0.67 & 21.84$\pm$0.112 &.....&  ....  \\
\hline
J0403-43:E5 &        04~04~34.0  & -43~24~53.1 &   518 &  2.8  &  41.5$\pm$1.65 & 21.24$\pm$0.068 &18.89$\pm$0.06& -0.16$\pm$0.10\\
J0403-43:E6 &        04~04~34.0  & -43~24~57.0 &   555 &  2.8  &  25.5$\pm$1.12 & 21.85$\pm$0.112 &.....&  ....\\
J0403-43:E7 &        04~04~33.9  & -43~24~49.5 &   182 &  2.8  &  12.2$\pm$0.75 & 21.44$\pm$0.080 &.....&  ....  \\
\hline
J0403-43:E8 &        04~04~34.4  & -43~21~05.2 &   493 &  2.4  &  23.4$\pm$1.05 & 21.81$\pm$0.109  &19.67$\pm$0.07& -0.12$\pm$0.11\\
J0403-43:E9 &        04~04~34.5  & -43~21~08.2 &   534 &  2.4  &  17.0$\pm$0.87 & 22.24$\pm$0.157  &.....&  ....\\
J0403-43:E10 &        04~04~33.6  & -43~20~56.0 &   215 &  2.4  &  12.9$\pm$0.77 & 21.56$\pm$0.088 &.....&  .... \\
\hline
J0403-43:E11 &        04~04~34.0  & -43~19~56.1 &   591 &  2.4  &  13.9$\pm$0.79 & 22.57$\pm$0.207  &20.24$\pm$0.07&-0.22$\pm$0.12\\
J0403-43:E12 &        04~04~33.9  & -43~19~49.9 &   387 &  2.4  &  5.86$\pm$0.64 & $>$23.05  &.....& .... \\
\hline
J0403-43:E13 &        04~04~34.3  & -43~19~20.6 &   202 &  2.5  &  7.27$\pm$0.66 & 22.11$\pm$0.141 &19.93$\pm$0.07&  -0.04$\pm$0.11\\
\hline
J0403-43:E14 &        04~04~30.3  & -43~19~16.7 &   425 &  2.2  &  7.39$\pm$0.66 & 22.90$\pm$0.272 &19.83$\pm$0.07& -0.07$\pm$0.11\\
J0403-43:E15 &        04~04~29.5  & -43~19~01.1 &   221 &  2.2  &  10.1$\pm$0.71 & 21.85$\pm$0.113  &.....&  .... \\
\hline
J0403-43:E16 &        04~03~18.1  & -43~17~42.7 &   271 &  2.5  & 6.90$\pm$0.65 & 22.49$\pm$0.193   &  21.37$\pm$0.07  &   -0.17$\pm$ 0.13\\ 
\hline
J0403-43:E17 &        04~03~18.4  & -43~17~33.6 &   241 &  2.5  & 7.71$\pm$0.66 & 22.24$\pm$0.157   &  21.35$\pm$ 0.07  &   -0.25 $\pm$0.14 \\
\hline
J0403-43:E18 &        04~03~23.8  & -43~16~10.1 &   713 &  2.5  & 10.8$\pm$0.72 & $>$23.05          &  22.90$\pm$ 0.12  &   -0.58$\pm$0.30\\
\hline
J0403-43:E19 &        04~04~05.0  & -43~14~27.9 &   188 &  2.2  &4.37$\pm$0.62 & 22.59$\pm$0.210   &  24.23$\pm$0.54   &   0.56 $\pm$0.66  \\  
\hline
J0403-43:E20 &        04~03~43.9  & -43~14~32.5 &   587 &  2.2  & 8.89$\pm$0.68 & $>$23.05 	    &  20.54$\pm$0.07  &    -0.01 $\pm$0.11\\
\hline
J0403-43:E21 &        04~03~36.6  & -43~15~07.3 &   360 &  2.2  & 5.46$\pm$0.63 & $>$23.05          &  20.97$\pm$0.07  &   -0.11 $\pm$0.14\\
\hline
J0403-43:E22 &        04~03~47.0  & -43~14~28.3 &   420 &  2.2  & 6.38$\pm$0.64 & $>$23.05  	    &  21.81$\pm$0.08  &   -0.14 $\pm$0.16\\
\hline
J0403-43:E23 &        04~03~45.8  & -43~14~46.3 &   887 &  2.1  & 13.4$\pm$0.78 & $>$23.05	    &  22.52$\pm$0.09  &   -0.34 $\pm$0.21\\
\hline
J0403-43:E24 &        04~03~46.9  & -43~14~41.8 &   415 &  2.1  &  10.1$\pm$0.71 & 22.54$\pm$0.202  &  21.71$\pm$0.08  &   -0.12 $\pm$0.15 \\

 \enddata
\tablecomments{Column descriptions [units]: (1) Source name.  (2) Source Right Ascension [J2000 hms] (3) Source declination [J2000 dms]. (4) Equivalent Width (5) Number of times beyond the 25$^{th}$ magnitude elliptical isophote the source lies (6) Emission-line flux of the source measured from SINGG image [10$^{-16}$ ergs s$^{-1}$ cm$^{-2}$] (7) R-band magnitude [AB mags]  (8) GALEX FUV magnitude (AB magnitude) of the source closest to the SINGG ELdot corrected for Galactic extinction (9) FUV - NUV color. Pertaining to columns 8 and 9: due to resolution discrepancies between SINGG images (1.5") and GALEX images (4.5") multiple ELdots are detected as single sources in the GALEX images. We have delineated affected sources by the horizontal lines in the table.  The GALEX image for J0403-43 comes from the Nearby Galaxy Atlas (NGA), and has FUV and NUV exposure times of 2370 seconds.  }
%\tablecomments{{\it Sample portion of table.}}
\end{deluxetable}
%%%%%%%%%%%%%%%%%%%%%%%%%%%%%%%%%%%%%%%
\begin{deluxetable}{l c c c c c c}
  \tablecaption{GALEX Ultraviolet Properties of ELdots \label{galextab}}
  \tabletypesize{\scriptsize}
%  \rotate
  \tablewidth{0pt}
  \tablehead{\colhead{ELdots with} &
             \colhead{Line Obs.} &
             \colhead{GI Program/Image ID} &
             \colhead{FUV Exptime} &
             \colhead{NUV Exptime} &
             \colhead{FUV} &
             \colhead{FUV - NUV} \\
             \colhead{Spectra} &
             \colhead{} &
             \colhead{} &
             \colhead{s} &
             \colhead{s} &
             \colhead{AB mag} &
             \colhead{AB mag}}
            
            \startdata
 J0005-28:E1 &  [OII]  &GI1\_009016\_HPJ0005m28*   & 1608 & 1608 &  $>$23.62 & 0.51$\pm$0.45\\
J0221-05:E2  & [OIII] &   XMMLSS\_03\_Meurer* &            25915.7    &3907    &  22.56$\pm$0.11 & 0.33$\pm$0.24\\
J0409-56:E1/2 & \ha  &   GI3\_087004\_NGC1533  &     1520&  3152 &   20.74$\pm$0.08& -0.14$\pm$0.13\\ 
J0409-56:E3  &  \ha      &  "  &      "       &"  &    22.67$\pm$0.14 &0.03$\pm$0.28\\
J0409-56:E4  &  \ha       &   "        &  "  &  "  & 22.35$\pm$0.12& -0.79$\pm$0.37 \\
J0506-31:E1 & [OIII] & NGA\_NGC1800\_000*1 & 1695 & 1695 & 23.38$\pm$0.13 & -0.30$\pm$0.34\\
J0507-37:E1 &  [OIII] & GI2\_121004\_LGG127*  &     2155&  2155 &   22.26$\pm$0.13 & 0.49$\pm$0.22\\
J0512-32:E1  & [OIII] & GI1\_047029\_UGCA106*    &  1637 & 4476   & 22.14$\pm$0.12& 0.05$\pm$0.23\\
J2149-60:E1 &  [OIII] & GI1\_009095\_NGC7125 *   &  1687 & 2664 &   22.55$\pm$0.15&-0.55$\pm$0.46\\
J2352-52:E1  & [OIII] & GI1\_047112\_ESO149\_G003 & 1565 & 1624   & 22.75$\pm$0.15 & 0.90$\pm$0.21\\
\hline
ELdots without\\
 Spectra \\
\hline
J0019-22:E1 &.....    & GI1\_047003\_MCG\_04\_02\_003* &3448 & 1664 &   20.37$\pm$0.07& 0.18$\pm$0.11\\
J0019-22:E2 &.....    &     "        & "&  " & 21.33$\pm$0.08  &0.10$\pm$0.17\\
J0031-22:E1 &.....   & GI1\_009007\_HPJ0031m22*&     1608 & 1608&   22.01$\pm$ 0.05 &     0.28$\pm$ 0.07\\
J0031-22:E2 &.....     &  "      &  "    & "   & 22.97$\pm$0.05 & 0.64$\pm$0.07\\
J0221-05:E1  & .....   & XMMLSS\_03\_Meurer*   &   25915.7 & 3907&    23.65$\pm$0.21 & 0.55$\pm$0.44\\
J0221-05:E3  & .....  &  "   &    "          &  "  & 23.19$\pm$0.16 & 0.63$\pm$0.31\\
J0224-24:E1 &.....    & GI1\_009115\_NGC0922* &      2285&  2285 &   22.53$\pm$0.12 & 0.27$\pm$0.24\\
J0320-52:E1 &.....   &  MIS2DFR\_38632\_0859    &   1855 & 1855 &   22.39$\pm$0.11& -0.40$\pm$0.26 \\
J0355-42:E1 &.....    & GI1\_047022\_NGC1487*    &   1696 & 1696  &  22.55$\pm$0.11 &-0.32$\pm$0.28\\
J0359-45:E1 &.....    & GI1\_047023\_HorDwarf *  &   3116 & 1510 &   23.57$\pm$0.24 & 0.66$\pm$0.37 \\

 \enddata
\tablecomments{Column descriptions: (1) Source name.  (2) The emission-line spectroscopically confirmed to fall within the SINGG narrow-band, if available. \ha~indicates star-formation associated with the SINGG galaxy; [OIII] falling within the \ha~ bandpass indicates a source redshift of approximately 0.3; and [OII] indicates a redshift of 0.76. (3) The GALEX Guest Investigator program identification that provided the GALEX images. An asterisk indicates that the field is part of the SUNGG sample. (4) and (5) NUV and FUV exposure times (6) Apparent FUV magnitude of the source at the position of the ELdot. In one case, J0409-56:E1 \& 2, two separate ELdots appear as a single source, due to the larger PSF of the GALEX images (4.6$\arcsec$) compared to the SINGG \ha images ($\sim1.5$ $\arcsec$) (7) FUV - NUV colors of the ELdots   }
%\tablecomments{{\it Sample portion of table.}}
\end{deluxetable}
%\begin{landscape}

%%%%%%%%%%%%%%%%%%%%%%%%%%%%%%%%%
\end{document}